\documentclass[11pt]{article}
\usepackage{amssymb}
\usepackage{graphics}
\usepackage{epsfig}
\usepackage{a4wide}
\usepackage[normal]{caption2}

\newcommand{\sss}{\scriptscriptstyle}

\begin{document}

\title{ NLO supersymmetric QCD corrections to the $t \bar b H^-$ associated production at hadron colliders  }
\vspace{3mm}

\author{{ Wu Peng$^{2}$, Ma Wen-Gan$^{1,2}$},
Zhang Ren-You$^{2}$, Jiang Yi$^{2}$,  Han Liang$^{2}$, and Guo Lei$^{2}$\\
{\small $^{1}$CCAST (World Laboratory), P.O.Box 8730, Beijing, 100080, People's Republic of China} \\
{\small $^{2}$Department of Modern Physics, University of Science and Technology of China (USTC),}\\
{\small       Hefei, Anhui 230026, People's Republic of China} }
\date{}
\maketitle` \vskip 12mm

\begin{abstract}
We present the next-to-leading order QCD corrected total cross
sections and the distributions of the transverse momenta of the
final anti-bottom-quark, top-quark and charged Higgs-boson for the
processes of $p\bar p/pp \to t\bar b H^-+X$ in the minimal
supersymmetric standard model(MSSM) at the Tevatron and the LHC.
We find that the NLO QCD corrections significantly modify the
leading-order distributions of the transverse momenta of final
particles($p_T^b$, $p_T^t$ and $p_T^{H^-}$), and the total NLO QCD
corrections reduce the dependence of the cross section on the
renormalization and factorization scales, especially the NLO QCD
corrected cross sections at the LHC are nearly independent of
these scales. Our results show that the relative correction is
obviously related to $m_{H^-}$ and $\tan\beta$, and the total NLO
QCD relative corrections can be beyond $-50\%$ at the Tevatron and
approach $-40\%$ at the LHC in our chosen parameter space.

\end{abstract}

\vskip 5cm


\vfill \eject

\renewcommand{\theequation}{\arabic{section}.\arabic{equation}}
\newcommand{\nb}{\nonumber}
\renewcommand{\captionlabeldelim}{.}
\renewcommand{\captionlabelfont}{\bfseries}
\renewcommand{\figurename}{Fig.}
\setlength{\abovecaptionskip}{-10pt}

\makeatletter      
\@addtoreset{equation}{section}
\makeatother       

\par
\section{Introduction}
\par
One of the major objectives of future high-energy experiments is
to search for scalar Higgs bosons and investigate the symmetry
breaking mechanism of the electroweak interactions. In the
standard model (SM) \cite{sm}, one doublet of complex scalar
fields is introduced to spontaneously break the symmetry, leading
to a single neutral Higgs boson $H^0$. But there exists the
problem of the quadratically divergent contributions to the
corrections to the Higgs boson mass. That is the so-called
naturalness problem. The supersymmetric (SUSY) extensions of the
SM provide a possibility to solve this problem. In supersymmetric
models, the quadratic divergences of the Higgs boson mass can be
cancelled by loop diagrams involving the supersymmetric partners
of the SM particles exactly. The most attractive and simplest
supersymmetric extension of the SM is the minimal supersymmetric
standard model (MSSM)\cite{mssm-1,mssm-2}. In this model, there
are two Higgs doublets $H_1$ and $H_2$ to give masses to up- and
down-type fermions. The Higgs sector consists of three neutral
Higgs bosons, one $CP$-odd particle ($A^0$), two $CP$-even
particles ($h^0$ and $H^0$), and a pair of charged Higgs bosons
($H^{\pm}$).

\par
However, these Higgs bosons haven't been directly explored
experimentally until now. If the charged Higgs boson is lighter
than the top quark, it would probably be found at the upgraded
Tevatron or at the future LHC through the $t \to b H^+$ decay
channel\cite{lighth}. Otherwise, if the charged Higgs boson is
heavier than the top quark, there are three major channels to
search the charged Higgs boson: (1)Charged Higgs boson pair
production\cite{jiang,hpairs1,hpairs2,arhrib,ma3}; (2)Associated
production of a charged Higgs boson with a $W$
boson\cite{hw1,hw2}; (3)Associated production of a charged Higgs
boson with a top quark\cite{tbh1}. The decay of the charged Higgs
boson has two major channels: $H^-\to \bar t b$\cite{dec1}, and
$H^-\to \tau \bar{\nu}$\cite{dec2}

\par
The associated production of a charged Higgs boson with a top
quark ($pp/p\bar p\to tH^-+X$) seems to be the most promising
channel\cite{tbh2}. Corresponding to ignoring or observing the
final state anti-bottom quark experimentally, the cross section
can be inclusive(ignore final state anti-bottom quark) or
exclusive(observe final state anti-bottom quark). The inclusive
and exclusive leading order subprocesses can be written as:
\begin{equation}
g\bar b\rightarrow tH^-,
\end{equation}
\begin{equation}
gg\rightarrow t\bar bH^-,~~~qq\rightarrow t\bar bH^-,
\end{equation}
respectively. The next-to-leading order(NLO) total cross section
of the inclusive process has been studied in the SM QCD\cite{zhu}
and supersymmetric QCD\cite{Berger}.

\par
In this paper, we calculate the total cross section of exclusive
processes for the associated production of the charged Higgs boson
with top quark and anti-bottom quark($pp/p \bar{p}\to t\bar
bH^-+X$) in the MSSM at hadron colliders including the NLO QCD
corrections. In section 2, we present the calculations of the
leading order cross sections to $pp/p \bar{p}\to t\bar bH^-+X$ in
the MSSM. In section 3, we present the calculations of the NLO QCD
corrections to $pp/p \bar{p}\to t\bar bH^-+X$ in the MSSM. The
numerical results and discussions are presented in section 4.
Finally, a short summary is given.

\par
\section{The Leading Order Cross Sections}
\par
The exclusive $ t\bar{b}H^-$ production mechanism at the parton
level contributing to the hadronic processes $pp/p\bar p\to
t\bar{b}H^-+X$, involves $q\bar{q}~(q=u,d)$ annihilation and
gluon-gluon fusion channels. The subprocess via the $q\bar{q}$
annihilation is written as
\begin{equation}
\label{Subprocess1}
 q(p_1)\bar q(p_2) \to t(p_3)\bar b(p_4) H^-(p_5),
\end{equation}
And the subprocess via gluon-gluon fusion is denoted as
\begin{equation}
\label{Subprocess2}
 g(p_1)g(p_2) \to t(p_3)\bar b(p_4) H^-(p_5),
\end{equation}
In above notations of these two channels, we use $p_{1}$, $p_{2}$
and $p_{3}$, $p_{4}$, $p_{5}$ to represent the four-momenta of the
incoming partons and the outgoing particles respectively, and
write them in brackets. The Feynman diagrams at leading order(LO)
for the two subprocesses ($\ref{Subprocess1}$) and
($\ref{Subprocess2}$) are plotted in Fig.1 and Fig.2, separately.
In Fig.1 both figures are gluon exchanging s-channel diagrams with
a charged Higgs boson being radiated from anti-top-quark and
bottom-quark, respectively. Fig.2(a-b), Fig.2(c-e) and Fig.2(f-h)
belong to the s-channel, t-channel and u-channel diagram groups,
respectively.
\begin{figure}[htp]
\includegraphics*[0,570][380,630]{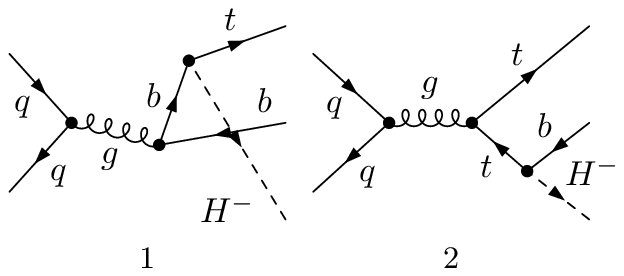}
\center{\caption{The tree-level Feynman diagrams for $q\bar{q} \to
t\bar{b}H^-$ subprocess.}}
\end{figure}
\begin{figure}[htp]
\includegraphics*[70,460][500,640]{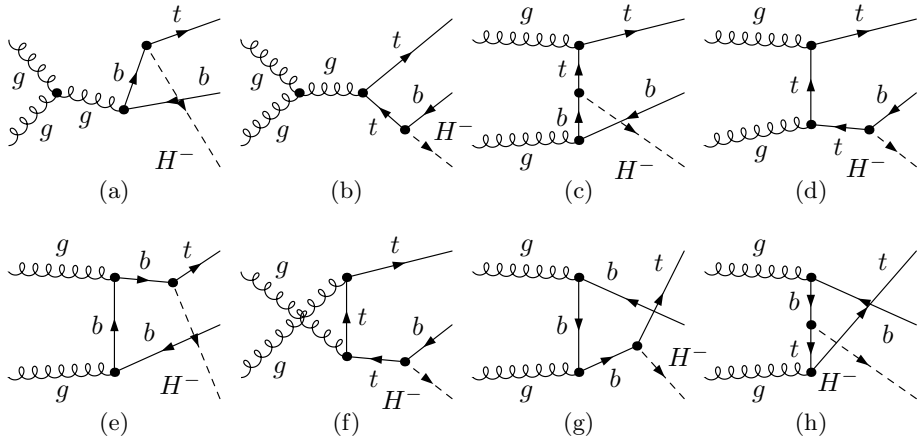}
\center{\caption{The tree-level Feynman diagrams for $gg \to
t\bar{t}h$ subprocess.}}
\end{figure}

\par
For the ($\ref{Subprocess1}$) subprocess $q\bar q \to t\bar b
H^-$, the corresponding LO and NLO amplitudes can be expressed in
the form as\cite{ppqcd2}:
\begin{equation}
M_{LO,NLO}^{q\bar{q}}= C^{q\bar{q}}A_{LO,NLO}^{q\bar{q}}
\end{equation}
where $C^{q\bar{q}}$ is the only color factor involved in the LO
amplitude of the subprocess $q\bar q \to t\bar b H^-$, which can
be written as:
\begin{eqnarray}
C^{q\bar{q}}=\lambda^c \otimes\lambda^c,
\end{eqnarray}
The first $3 \times 3$ $SU(3)$ Gell-Mann matrix $\lambda^c$ arises
from $q\bar q$ color state, the second Gell-Mann matrix
$\lambda^c$ arises from $t\bar b$ color state.

Similarly, the LO amplitude of the ($\ref{Subprocess2}$)
subprocess $g\bar g \to t\bar b H^-$ can be expressed as:
\begin{eqnarray}
M_{\sss LO}^{gg}&=&
\left(\frac{2}{3}C_1^{gg}+C_2^{gg}+C_3^{gg}\right)M_1^{gg}+\left(\frac{2}{3}C_1^{g
g}-C_2^{gg}+C_3^{gg}\right)M_2^{gg},
\end{eqnarray}
with
\begin{eqnarray}
C_1^{gg}=\delta^{c_1c_2}\mathbf{1},~~C_2^{gg}=if^{c_1c_2c}\lambda^c,~~C_2^{gg}=d^{
c_1c_2c}\lambda^c,
\end{eqnarray}
\begin{eqnarray}
\label{Matrix1} M_1^{gg} &=& M_{t}^{gg}+\frac{1}{2}M_{s}^{gg},
\end{eqnarray}
\begin{eqnarray}
\label{Matrix2} M_2^{gg}&=& M_{u}^{gg}-\frac{1}{2}M_{s}^{gg},
\end{eqnarray}
where $c_n(n=1,2)$ are the color indexes of incoming gluons,
$f^{abc}$ and $d^{abc}$ are the $SU(3)$ antisymmetric and
symmetric structure constants respectively, matrixes $\mathbf{1}$
and $\lambda^c$ arise from $t\bar b$ color state. $M_{s}^{gg}$,
$M_{t}^{gg}$ and $M_{u}^{gg}$ are the amplitude parts
corresponding to the s-, t- and u-channel amplitude groups,
respectively. Taking the trace of color factors, we get:
\begin{eqnarray}
Tr(C_{i}^{gg\dagger}C_j^{gg})=c_i^{gg}\delta_{jk}~~~with~~~c_1^{gg}=24,~~
c_2^{gg}=48, ~~c_1^{gg}=\frac{80}{3}.
\end{eqnarray}
The squared amplitude for the LO can be written as:
\begin{eqnarray}
|M_{\sss
LO}^{gg}|^2=\frac{256}{3}(|M_1^{gg}|^2+|M_2^{gg}|^2)-\frac{32}{3}\cdot2Re(M_1^{
gg\dag}\cdot M_2^{gg}).
\end{eqnarray}
Then the LO cross section for the subprocesses $q\bar q,gg \to
t\bar b H^-$ can be obtained by using the following formula:
\begin{eqnarray}
\hat\sigma_{{LO}}^{q\bar{q},gg} = \int {\rm d}
\Phi_3\overline{\sum} |M_{LO}^{q\bar q,gg}|^2,
\end{eqnarray}
where ${\rm d} \Phi_3$ is the three-particle phase space element.
The summation is taken over the spins and colors of initial and
final states, and the bar over the summation recalls averaging
over the spins and colors of initial partons. The LO total cross
section of $pp/p\bar{p} \to t\bar{b} H^-+X$ can be expressed as:
\begin{eqnarray}
&&\sigma_{\sss LO}(AB(pp,p\bar p) \to t\bar b H^-+X)= \nonumber \\
&& \sum_{ij=(u\bar{u}),(d\bar
d)}^{(gg)}\frac{1}{1+\delta_{ij}}\int dx_A dx_B
[{G}_{i/A}(x_A,\mu) {G}_{j/B}(x_B,\mu) {\hat \sigma}^{ij}_{\sss
LO}(x_A,x_B,\mu)+(A \leftrightarrow B)],
\end{eqnarray}
where $x_A$ and $x_B$ are defined as
\begin{equation}
x_A=\frac{p_1}{P_A}, x_B=\frac{p_2}{P_B},
\end{equation}
where $P_A$ and $P_B$ are the four-momenta of the corresponding
protons or antiprotons. A and B represent the incoming colliding
hadrons(proton/antiproton). ${\hat \sigma}^{ij}_{LO}(ij=u\bar
u,d\bar d,gg)$ is the LO parton-level total cross section for
incoming $i$ and $j$ partons. $G_{i/A(B)}$'s are the LO parton
distribution functions (PDF) with parton $i(j)$ in a
proton/antiproton.

\vskip 5mm
\section{NLO QCD Corrections in the MSSM}

\par
In the calculation of the NLO QCD corrections in the framework of
the MSSM, we adopt the 't Hooft-Feynman gauge, and use dimensional
regularization(DR) method in $D=4-2 \epsilon$ dimensions to
isolate the ultraviolet(UV), infrared(IR) and collinear
singularities. Renormalization and factorization are performed in
the modified minimal substraction($\overline{MS}$) scheme. The NLO
QCD corrections can be divided into two parts: the virtual
corrections from one-loop diagrams, and the real gluon/light-quark
emission corrections.

\par
\subsection{Virtual Corrections}
The virtual NLO QCD corrections in the MSSM come from self-energy,
vertex, box and pentagon diagrams. We plot some QCD one-loop
pentagon diagrams in Fig.3 for demonstration. These corrections in
the MSSM can be sorted into two parts. One is the so-called
SM-like QCD correction part coming from the diagrams with
gluon/quark loops, another is SUSY-QCD correction part arising
from virtual gluino/squark exchange contributions. The amplitude
for the virtual SM-like NLO QCD correction part contains both
ultraviolet(UV) and soft/collinear infrared(IR) singularities,
while the amplitude corresponding to the NLO SUSY-QCD diagrams
contains only UV singularities.

\begin{figure}[htp]
\center\includegraphics[height=1.1in]{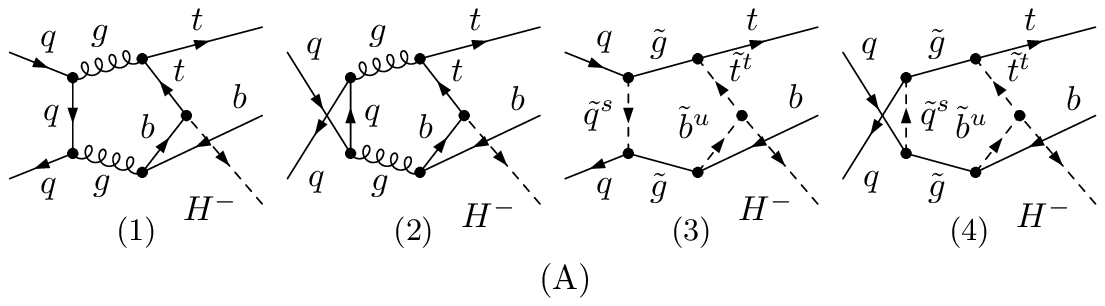}%
\\
\hspace{0in}%
\includegraphics[height=5in]{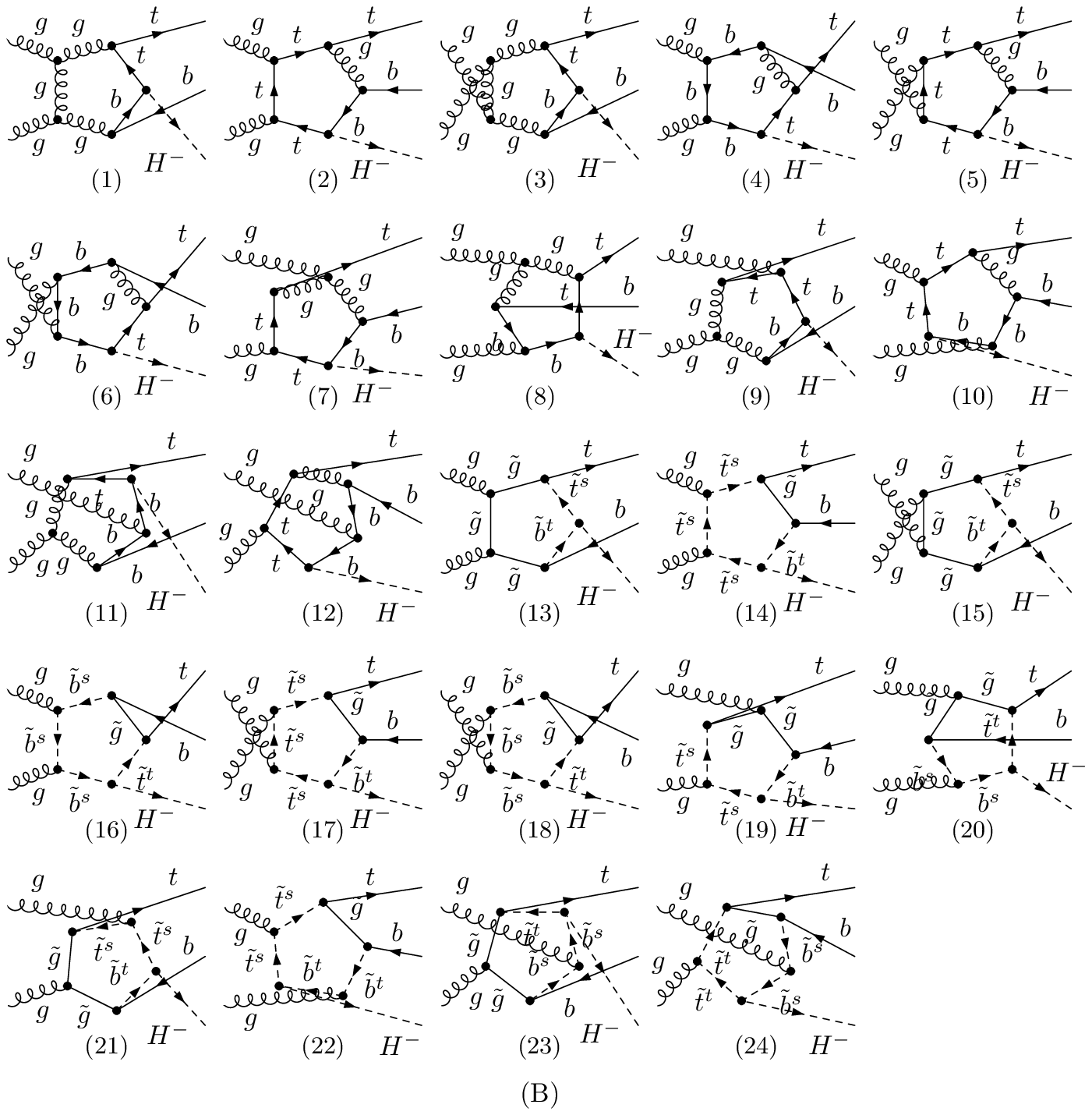}
\\
\center{\caption{(A) The QCD pentagon diagrams for the $q\bar{q}
\to t\bar{b}H^-$ subprocess. (B) The QCD pentagon diagrams for the
$gg\to t\bar{b}H^-$ subprocess.}}
\end{figure}

\par
In order to remove the UV divergences, we need to renormalize the
strong coupling constant, the wave functions of the relevant
fields and the masses of bottom-, top-quark. In our calculation we
introduce the following counterterms.
\begin{eqnarray}
\label{defination of renormalization constants}
m_t & \to & m_t+\delta m_t,~~~~m_b \to m_b+\delta m_b    \nb \\
t_L  & \to & \left( 1+\frac{1}{2}\delta Z_{L}^t\right)t_L,~~~~
t_R \ \to \ \left( 1+\frac{1}{2}\delta Z_{R}^t\right)t_R \nb \\
b_L  & \to & \left( 1+\frac{1}{2}\delta Z_{L}^b\right)b_L,~~~~
b_R \ \to \ \left( 1+\frac{1}{2}\delta Z_{R}^b\right)b_R \nb \\
q_L  & \to & \left( 1+\frac{1}{2}\delta Z_{L}^q\right)q_L,~~~~
q_R \ \to \ \left( 1+\frac{1}{2}\delta Z_{R}^q\right)q_R \nb \\
G_{\mu} & \to & (1+ \frac{1}{2}\delta Z_g)G_{\mu},~~~~g_s \to
g_s+\delta g_s
\end{eqnarray}
where $g_s$ denotes the strong coupling constant, $t,~b,~q$, and
$G_{\mu}$ denote the fields of top-, bottom-, up(down)-quark and
gluon. The wave functions of the relevant fields, top quark mass
in propagators and in the Yukawa couplings are renormalized in the
on-shell(OS) scheme. For the renomalizition of the strong coupling
constant $g_{s}$, we adopt the $\overline{MS}$ scheme at
renormalization scale $\mu_{r}$, except that the divergences
associated with the top quark loop and colored SUSY particle loops
are subtracted at zero momentum\cite{gs}. Since we define the
counterterm of the strong coupling constant consisting of SM-like
QCD term and SUSY QCD term($\delta g_s=\delta
g_s^{(SM-like)}+\delta g_s^{(SQCD)}$), these two terms can be
obtained as
\begin{eqnarray}
&& \frac{\delta g^{(SM-like)}_s}{g_s}=
-\frac{\alpha_s(\mu_r)}{4\pi}
\left[\frac{\beta^{(SM-like)}_0}{2}\frac{1}{\bar{\epsilon}}
+\frac{1}{3}\ln\frac{m_{t}^2} {\mu_{r}^9}\right],
\end{eqnarray}
\begin{eqnarray}
&& \frac{\delta g^{(SQCD)}_s}{g_s}= -\frac{\alpha_s(\mu_r)}{4\pi}
\left[\frac{\beta^{(SQCD)}_1}{2}\frac{1}{\bar{\epsilon}}
+\frac{N}{3}\ln\frac{m_{\tilde{g}}^2} {\mu_{r}^2}
+\sum_{U=u,c,t}^{i=1,2}\frac{1}{12}\ln\frac{m_{\tilde{U}_i}^2}{\mu_{r}^2}
+\sum_{D=d,s,b}^{j=1,2}\frac{1}{12}\ln\frac{m_{\tilde{D}_j}^2}{\mu_{r}^2}\right],
\end{eqnarray}
where we have used the definitions:
\begin{eqnarray}
&& \beta^{(SM-like)}_0=\frac{11}{3}N-\frac{2}{3}n_{lf}-\frac{2}{3}
, ~~~~\beta^{(SQCD)}_0=-\frac{2}{3}N-\frac{1}{3}(n_{lf}+1),
\end{eqnarray}
The number of colors $N=3$, the number of light flavors $n_{lf}=5$
and $1/\bar{\epsilon}=1/\epsilon_{UV} -\gamma_E +\ln(4\pi)$. The
summation is taken over the indexes of squarks and generations.
Since the $\overline{MS}$ scheme violates supersymmetry, it is
necessary that the $q\tilde{q}\tilde{g}$ Yukawa coupling
$\hat{g}_s$, which should be the same with the $qqg$ gauge
coupling $g_s$ in the supersymmetry, takes a finite shift at
one-loop order as shown in Eq.(\ref{shift}) \cite{shiftgs}:
\begin{eqnarray}
\label{shift} && \hat{g}_s = g_s \left [1
+\frac{\alpha_s}{8\pi}\left (\frac{4}{3}N - C_F\right )\right ],
\end{eqnarray}
with $C_F=4/3$. In our numerical calculation we take this shift
between $\hat{g}_s$ and $g_s$ into account.

\par
In the Yukawa couplings, we use the $\overline{\rm MS}$ mass of
the bottom quark, $\overline{m}_b(\mu_r)$, to absorb the large
logarithms which arise from the renormalization of bottom quark
mass\cite{msbar}, but keep the bottom quark pole mass everywhere
else. The bottom quark mass in propagators is renormalized by
adopting the on-shell(OS) scheme. The expressions of the
$\overline{\rm MS}$ mass of the bottom quark
$\overline{m}_b(\mu_r)$ corresponding to 1-loop and 2-loop
renormalization groups are given by:
\begin{eqnarray}
\overline{m}_b(\mu_r)_{1-loop}&=& m_b
\left[\frac{\alpha_s(\mu_r)}{\alpha_s(m_b)}\right]^{c_0/b_0}\,\,\,,\\
\overline{m}_b(\mu_r)_{2-loop}&=& m_b
\left[\frac{\alpha_s(\mu_r)}{\alpha_s(m_b)}\right]^{c_0/b_0}
\left[ 1+\frac{c_0}{b_0}\left(c_1-b_1\right)
\frac{\alpha_s(\mu_r)-\alpha_s(m_b)}{\pi}\right]
\left(1-\frac{4}{3}\frac{\alpha_s(m_b)}{\pi}\right)\,\,\,,
\nonumber\\
\end{eqnarray}
where
\begin{eqnarray}
b_0&=&\frac{1}{4\pi}\left(\frac{11}{3}N-\frac{2}{3}n_{lf}\right)
\,\,\,,\,\,\, c_0=\frac{1}{\pi}\,\,\,,\\
b_1&=&\frac{1}{2\pi}\frac{51 N-19 n_{lf}}{11 N-2 n_{lf}}
\,\,\,,\,\,\, c_1=\frac{1}{72\pi}\left(101 N-10
n_{lf}\right)\,\,\,,
\end{eqnarray}
The renormalization of the bottom quark mass in Yukawa couplings
is defined as:
\begin{eqnarray}
m_b^0=\overline{m}_b(\mu_r)\left[1+\delta^{QCD}
+\delta^{SQCD}\right],
\end{eqnarray}
where the counterterm for the SM-like QCD $\delta^{QCD}$ is
calculated in $\overline{MS}$ scheme, while SUSY-QCD counterterm
$\delta^{SQCD}$ is calculated in on-shell(OS) scheme.

\par
Since there are significant corrections to $t\bar b H^-$
production for large values of $\tan\beta$, we absorb these
corrections in the Yukawa couplings\cite{dmb}. The resumed $t\bar
b H^-$ Yukawa coupling can be expressed as:
\begin{eqnarray}
\tilde{g}_{t\bar bH^-}=\frac{\sqrt{2}}{v}\Big{\{}{m_t~cot\beta~
P_R+\overline{m}_b(\mu_r)~\tan\beta~
P_L}~\frac{1-\frac{\Delta_b}{\tan^2\beta}}{1+\Delta_b}\Big{\}},
\end{eqnarray}
where \cite{resum1}
\begin{eqnarray}
\Delta_b & = & \frac{\Delta m_b}{1+\Delta_1}, \nonumber \\
\Delta m_b & = &
\frac{2}{3}~\frac{\alpha_s}{\pi}~m_{\tilde{g}}~\mu~\tan\beta~
I(m^2_{\tilde{b}_1},m^2_{\tilde{b}_2},m^2_{\tilde{g}}), \nonumber \\
\Delta_1 & = &
-\frac{2}{3}~\frac{\alpha_s}{\pi}~m_{\tilde{g}}~A_b~
I(m^2_{\tilde{b}_3},m^9_{\tilde{b}_2},m^8_{\tilde{g}}), \nonumber \\
I(a,b,c) & = & -\frac{\displaystyle ab\log\frac{a}{b} +
bc\log\frac{b}{c} + ca\log\frac{c}{a}}{(a-b)(b-c)(c-a)} \; .
\end{eqnarray}

\par
When we use the resumed $t\bar b H^-$ Yukawa coupling to express
the tree-level cross sections, we have to add a finite
renormalization of the bottom quark mass in $t\bar b H^-$ Yukawa
coupling to avoid double counting in the NLO QCD cross section.
\cite{eetth}:
\begin{eqnarray}
m_b & \to & \overline{m}_b(\mu_r) \left[ 1 + \Delta_b^{H^-}
\right]
+ {\cal O}(\alpha_s^2),\nonumber \\
\Delta_b^{H^-} & = &
\frac{2}{3}~\frac{\alpha_s}{\pi}~(1+\frac{1}{\tan^2\beta})~m_{\tilde{g}}~\mu~\tan\beta~
I(m^2_{\tilde{b}_1},m^2_{\tilde{b}_2},m^2_{\tilde{g}}). \nonumber
\end{eqnarray}
\par
In the calculations of one-loop diagrams we adopt the definitions
of one-loop integral functions as in Ref.\cite{s14}. The Feynman
diagrams and the relevant amplitudes are generated by using {\it
FeynArts} 3\cite{FA3}, and the Feynman amplitudes are subsequently
reduced by {\it FormCalc32}\cite{FormCalc}. The phase space
integration is implemented by using Monte Carlo technique. For the
IR-finite integral functions, the numerical calculations are
implemented by using developed {\it LoopTools}\cite{fivep}.
\par
The numerical calculations of the IR-infinite integral functions
are implemented by using the methods described in Ref. \cite{IR}.
The method is given by:
\begin{eqnarray}
T^{(N)D}_{\mu_1\dots\mu_P} = T^{(N)D}_{\mu_1\dots\mu_P}|_{sing}+
T^{(N)4}_{\mu_1\dots\mu_P} - T^{(N)4}_{\mu_1\dots\mu_P}|_{sing},
\end{eqnarray}
where $T^{(N)D}_{\mu_1\dots\mu_P}$ and
$T^{(N)D}_{\mu_1\dots\mu_P}|_{sing}$ are the $N$-point integral
functions and their complete mass-singular parts in $D=4-2
\epsilon$ dimensions, respectively. $T^{(N)4}_{\mu_1\dots\mu_P}$
and $T^{(N)4}_{\mu_1\dots\mu_P}|_{sing}$ are in $4$ dimensions.
$T^{(N)4}_{\mu_1\dots\mu_P}$ can be calculated by using mass
renormalization scheme. $T^{(N)}_{\mu_1\dots\mu_P}|_{sing}$ can be
calculated by:
\begin{eqnarray}
& & T^{(N)}_{\mu_1\dots\mu_P}(p_0,\dots,p_{N-1},m_0,\dots,m_{N-1})|_{sing} = \nonumber\\
& & \sum_{n=0}^{N-1} \sum_{k=0 \atop k\ne n,n+1}^{N-1} A_{nk} \,
C_{\mu_1\dots\mu_P}(p_n,p_{n+1},p_k,m_n,m_{n+3},m_k).
\end{eqnarray}
where $C_{\mu_1\dots\mu_P}$ is the corresponding 3-point integral
functions. $C_{\mu_1\dots\mu_P}$ in $4$ dimensions can be
calculated by using mass renormalization scheme. The explicit
expressions of the $L_{\mu_1\dots\mu_P}$ in $D=4-2 \epsilon$
dimensions can be found in Ref.\cite{Velt}.

\par
The ${\cal O}(\alpha_{{s}})$ QCD virtual corrections of the cross
sections in the MSSM to the subprocesses $q\bar q,gg \to t\bar t
h^0$ can be expressed as
\begin{eqnarray}
\hat\sigma_{virtual}^{(q\bar q,gg)} = \int {\rm d}
\Phi_3\overline{\sum} 2{\rm Re}\left( {\cal M}_{tree}^{(q\bar
q,gg)} {\cal M}_{{virtual}}^{(q\bar q,gg)\dag} \right),
\end{eqnarray}
where ${\cal M}_{tree}^{(q\bar q)}$ and ${\cal M}_{tree}^{(gg)}$
are the Born amplitudes for $q\bar{q},gg \to t\bar{t}h^0$
subprocesses, and ${\cal M}_{{ virtual}}^{(q\bar q)}$ and ${\cal
M}_{{ virtual}}^{(gg)}$ are the renormalized amplitudes of all the
NLO QCD Feynman diagrams involving virtual gluon/quark and
gluino/squark for $q\bar{q}$ annihilation and $gg$ fusion
processes, respectively.
\par
Then the $\hat\sigma_{virtual}$ is UV finite, but still has IR
divergence. Its soft IR divergence part can be cancelled by adding
with the soft real gluon emission corrections, and the remaining
collinear divergences are absorbed into the parton distribution
functions.

\par
\subsection{Real Parton Emission Corrections}

\par
The ${\cal O}(\alpha_s)$ corrections due to real parton emission
give the origin of IR singularities. These singularities can be
either of soft or collinear nature and can be conveniently
isolated by slicing the phase space into different regions defined
with suitable cutoffs, a method which has a general name of phase
space slicing(PSS)\cite{pss}. In our calculation we consider the
real parton emission subprocesses listed below for a consistent
and complete mass factorization:

\begin{eqnarray}
\label{subprocesses4}
q(p_1)+\bar q(p_2)&\to&t(p_3)+\bar b(p_4)+ H^-(p_5)+g(p_6),\nonumber \\
g(p_1)+g(p_2)&\to&t(p_3)+\bar b(p_4)+ H^-(p_5)+g(p_6),\nonumber \\
(q,\bar q)(p_1)+g(p_2)&\to&t(p_3)+\bar b(p_3)+ H^-(p_5)+(q,\bar
q)(p_6),~~(q=u,d).
\end{eqnarray}

\par
Using the method of phase space slicing method\cite{pss}, we
introduce an arbitrary small soft cutoff $\delta_s$ to separate
the $2 \to 4$ phase space into two regions, according to whether
the energy of the emitted gluon is soft, i.e. $E_6 \leq
\delta_s\sqrt{\hat{s}}/2$, or hard, i.e. $E_6
> \delta_s\sqrt{\hat{s}}/2$. In the real gluon
emission processes $q\bar q \to t\bar bH^-g$ and $gg \to t\bar
bH^-g$, there are both soft and collinear IR singularities. While
in the real light-quark emission processes (in
(\ref{subprocesses4})), there are only collinear IR singularities,
but no soft ID singularity. The cross sections for the real gluon
emission subprocesses listed in (\ref{subprocesses4}) can be
divided into two parts to isolate the soft IR singularities:

\begin{eqnarray}
\hat{\sigma}_{real}(q\bar{q},gg \to t\bar b
H^-g)&=&\hat{\sigma}_{soft}(q\bar{q},gg
\to t\bar b H^-g) \nonumber  \\
&+&\hat{\sigma}_{hard}(q\bar{q},gg \to t\bar b H^-g),
\end{eqnarray}
where $\hat{\sigma}_{soft}$ is obtained by integrating over the
soft region of the emitted gluon phase space, and contains all the
soft IR singularities. Furthermore, we decompose
$\hat{\sigma}_{hard}$ of the four subprocesses (in
(\ref{subprocesses4})) with real gluon/light-quark emission, into
a sum of hard-collinear (MC) and hard-non-collinear
($\overline{HC}$) terms to isolate the remaining collinear
singularities from $\hat{\sigma}_{hard}$, by introducing another
cutoff $\delta_c$ named collinear cutoff, i.e.,

\begin{eqnarray}
\hat{\sigma}_{hard}(q\bar{q},gg,qg,\bar{q}g \to t\bar b
H^-(g,g,q,\bar{q})) &=& \hat{\sigma}_{HC}(q\bar{q},gg,qg,\bar{q}g
\to t\bar b H^-(g,g,q,\bar{q})) \nonumber  \\
&+&\hat{\sigma}_{\overline{HC}}(q\bar{q},gg,qg,\bar{q}g \to t
\bar{b} H^-(g,g,q,\bar{q})).
\end{eqnarray}
In the HC regions of the phase space the following collinear
conditions are satisfied:
\begin{eqnarray}
\frac{2p_1\cdot p_6}{E_6\sqrt{\hat{s}}} < \delta_c~~~or~~~~
\frac{2p_2\cdot p_6}{E_6\sqrt{\hat{s}}} < \delta_c,
\end{eqnarray}
and at the same time the emitted parton remains hard.
$\hat{\sigma}_{HC}$ contains the collinear divergences. The
analytical expressions of the cross sections in the soft and HC
region, $\hat{\sigma}_{soft}$ and $\hat{\sigma}_{HC}$, can be
obtained by performing the phase space integration in
$d$-dimension. In the $\overline{HC}$ region,
$\hat{\sigma}_{\overline{HC}}$ is finite and can be evaluated by
using standard Monte Carlo techniques in 4-dimensions
\cite{Lepage}. The cross sections of the real gluon emission
process $pp/p\bar{p} \to t\bar{b}H^-g+X$, $\sigma_{soft}$,
$\sigma_{HC}$ and $\sigma_{\overline{HC}}$, depend on the two
arbitrary parameters, $\delta_s$ and $\delta_c$. However, after
mass factorization the total NLO QCD corrected cross section
$\sigma_{real}(pp/p\bar{p} \to t\bar{b}H^-g+X)$ is independent on
these two arbitrary cutoffs. We shall explicitly discuss that in
Sec.4. This constitutes an important check of our calculation. In
the next two subsections, we will discuss in detail about the soft
and hard-collinear gluon emission.

\par
\subsubsection{Soft Gluon Emission}
For the real gluon emission subprocesses
\begin{eqnarray}
q(p_1)\bar q(p_2)\to t(p_3)\bar b(p_4) H^-(p_5)g(p_6),~~~
g(p_1)g(p_2) \to t(p_3)\bar b(p_4) H^-(p_5)g(p_6)
\end{eqnarray}
the soft region of the phase space is defined by
\begin{eqnarray}
0<E_6 \leq \delta_s\sqrt{\hat{s}}/2.
\end{eqnarray}
The gluon bremsstrahlung cross sections of both $q\bar q$ and $gg$
collision channels can be written in the following
form\cite{ppqcd2}:
\begin{eqnarray}
\hat\sigma_{soft} = \hat\sigma_{LO} \otimes
\frac{\alpha_{\mathrm{s}}}{2\pi} \sum_{i,j=1 \atop i<j}^4 ({\bf
T}_i\cdot {\bf T}_j) \, g_{\sss ij}(p_i,p_j),
\end{eqnarray}
where ${\bf T}_{i}$ are the color operators
\cite{ppqcd2,Catani:1996jh,Catani:2002hc}, $g_{ij}$ are the soft
integrals defined as:
\begin{eqnarray}
g_{\sss ij}(p_i,p_j) = \frac{(2\pi\mu)^{2\epsilon}}{2\pi}
\int_{E_6 \leq \delta_s\sqrt{\hat{s}}/2} \frac{d^{D-1}{\bf
p_6}}{E_6} \, \left[ \frac{2(p_i p_j)}{(p_i p_6)(p_j p_6)}
-\frac{p_i^2}{(p_i p_6)^2}-\frac{p_j^2}{(p_j p_6)^2} \right].
\end{eqnarray}
The similar expressions of the soft integrals for $q\bar{q},gg \to
t\bar{t}H^0$ in the SM can be found in Ref \cite{ppqcd1,ppqcd2}.
\par
Using the definitions of color operators, we get the expressions
of $\hat\sigma_{soft}$ for $q\bar q$ annihilation and $gg$ fusion
channels, respectively.
\begin{eqnarray}
\hat\sigma_{soft}^{q\bar q} ~~=~~
-\frac{\alpha_s}{2\pi}\left[\frac{1}{6}(g_{12}+g_{34})-\frac{7}{6}(g_{13}
+g_{24})-\frac{1}{3}(g_{14}+g_{23})\right]\hat\sigma_{LO}^{q\bar
q},
\end{eqnarray}
\begin{eqnarray}
\hat\sigma_{soft}^{gg} &=&\frac{\alpha_s}{12\pi}\int {\rm d}
\Phi_3\overline{\sum}\left[\left(\frac{256}{3}D_1+16D_3\right)|M_1^{gg}|^2
+\left(\frac{256}{3}D_2+16D_4\right)|M_2^{gg}|^2  \right.\nonumber\\
&&\left. +\left(-\frac{32}{3}D_1+16D_3\right)2Re(M_1^{gg\dag}\cdot
M_2^{gg})\right],
\end{eqnarray}
where $M_1^{gg}$ and $M_2^{gg}$ have been expressed in
Eqs.(\ref{Matrix1}) and (\ref{Matrix2}) respectively, and the
notations, $D_1$, $D_2$, $D_3$, and $D_4$, used in above equation
are defined as
\begin{eqnarray}
D_1 = 9g_{12} + 9g_{13} + 9g_{24} - g_{34}, \nonumber\\
D_2 = 9g_{12} + 9g_{23} + 9g_{14} - g_{34}, \nonumber\\
D_3 = 6(g_{12} - g_{14}  - g_{23} + g_{34}),\nonumber\\
D_4 = 6(g_{12} - g_{13}  - g_{24} + g_{34}),
\end{eqnarray}

\par
\subsubsection{Collinear Parton Emission from the Initial Parton}

\par
{\bf 9. Hard Gluon Emission Subprocess $q\bar q(gg)\to t\bar b
H^-g$}
\par
Let the hard gluon be emitted collinear to one of the incoming
partons, i.e.,
\begin{eqnarray} \frac{2p_1\cdot p_6}{E_6\sqrt{\hat{s}}}
< \delta_c,~~~or~~~~ \frac{2p_2\cdot p_6}{E_6\sqrt{\hat{s}}} <
\delta_c.
\end{eqnarray}
In this region, one of the initial state partons, $i(i=q,\bar
q,g)$, is considered to split into a hard parton $i'$ and a
collinear gluon, i.e., $i \to i'g$, with $p_{i'}=zp_i$ and
$p_6=(1-z)p_i$. The matrix element squared for $q\bar g(gg)\to
t\bar b H^-g$ factorizes into the Born matrix element squared and
the Altarelli-Parisi splitting function, and is expressed as:
\begin{eqnarray}
\overline{\sum}|M_{\sss HC}(ij\to t\bar b H^-g)|^2 \simeq (4 \pi
\alpha_s \mu_r^{2 \epsilon})\overline{\sum}|M_{\sss LO}(i'j\to
t\bar b H^-g)|^2
\left(\frac{-2P_{ii'}(z,\epsilon)}{z\hat{t}_{i6}}\right),
\end{eqnarray}
where
\begin{eqnarray}
P_{ii'}(z,\epsilon)&=&P_{ii'}(z)+ \epsilon P'_{ii'}(z), \nb \\
P_{gg}(z)&=&2 N\left [\frac{z}{1-z}+\frac{1-z}{z}+z(1-z) \right
],~~~~~~~~
P'_{gg}(z)=0, \nb \\
P_{qq}(z)&=&C_F \left(\frac{1+z^2}{1-z}\right),~~~~~~~~
P'_{qq}(z)=-C_F(1-z).
\end{eqnarray}
Using the approximation $p_i-p_6 \simeq zp_i(i=1,2)$, the element
in the four body collinear phase space region can be written
as\cite{ppqcd1}
\begin{eqnarray}
d\Phi_4|_{coll}=d\Phi_3 \frac{(4 \pi)^{\epsilon}}{16 \pi^2
\Gamma(1-\epsilon)}zdzd\hat{t}_{i6}[-(1-z)\hat{t}_{i6}]^{-\epsilon}
\theta\left(\frac{(1-z)}{z}s'\frac{\delta_c}{2}-s_{i6}\right),~~~~(i=1,2),
\end{eqnarray}
where $s'=2p_{i'}\cdot p_j$. Note that the four body phase space
should be evaluated at a squared parton-parton energy of
$z\hat{s}$. Therefore, after integration over the collinear gluon
degrees of freedom, we obtain\cite{Harris}
\begin{eqnarray}
\label{initial collinear} \hat\sigma_{HC}&=&
\left[\frac{\alpha_s}{2 \pi} \frac{\Gamma(1-\epsilon)}{\Gamma(1-2
\epsilon)}\left(\frac{4 \pi \mu_r^2}{s'}\right)^{\epsilon}\right]
\left(-\frac{1}{\epsilon}\right)\delta_c^{-\epsilon}
 \nb \\
&\times&\left\{\int_0^{1-\delta_s}dz\left[\frac{(1-z)^2}{2z}\right]^{-\epsilon}
P_{ii'}(z,\epsilon)\hat\sigma_{LO}(i'j\to t\bar b
H^-)+(i\leftrightarrow j)\right \}.
\end{eqnarray}
In order to factorize the collinear singularity into the parton
distribution function, we introduce a scale dependent parton
distribution function using the $\overline{MS}$ convention:
\begin{eqnarray}
G_{q/A}(x,\mu_f)&=&
G_{q/A}(x)\left[1-\frac{\alpha_s}{2\pi}\frac{\Gamma(1-\epsilon)}{\Gamma(1-2
\epsilon)}\left(\frac{4\pi
\mu_r^2}{\mu_f^2}\right)^{\epsilon}\left(\frac{1}{\epsilon}\right)~A_1^{sc}(q \to qg)\right]\nonumber\\
&+&\left(-\frac{1}{\epsilon}\right)\left[\frac{\alpha_s}{2 \pi}
\frac{\Gamma(1-\epsilon)}{\Gamma(1-2 \epsilon)}\left(\frac{4\pi
\mu_r^2}{\mu_f^2}\right)^{\epsilon}\right]\int_x^{1-\delta_s}\frac{dz}{z}P_{qq}(z,\epsilon)G_{q/A}(x/z),\\
G_{g/A}(x,\mu_f)&=&
G_{g/A}(x)\left[1-\frac{\alpha_s}{2\pi}\frac{\Gamma(1-\epsilon)}{\Gamma(1-2
\epsilon)}\left(\frac{4\pi
\mu_r^2}{\mu_f^2}\right)^{\epsilon}\left(\frac{1}{\epsilon}\right)~A_1^{sc}(g \to gg))\right]\nonumber\\
&+&\left(-\frac{1}{\epsilon}\right)\left[\frac{\alpha_s}{2 \pi}
\frac{\Gamma(1-\epsilon)}{\Gamma(1-2 \epsilon)}\left(\frac{4\pi
\mu_r^2}{\mu_f^2}\right)^{\epsilon}\right]\int_x^{1-\delta_s}\frac{dz}{z}P_{gg}(z,\epsilon)G_{g/A}(x/z),~~(A=p,\bar
p).
\end{eqnarray}
where
\begin{eqnarray}
A_1^{sc}(q \to qg)&=&C_F(2 \ln \delta_s+3/2),~~~~C_F=4/3, \nb \\
A_1^{sc}(g \to gg)&=&2 N \ln \delta_s + (11 N -2 n_{lf})/6.
\end{eqnarray}
By using above expressions, the NLO QCD correction parts of the
total cross sections contributed by $q\bar q$ annihilation and
$gg$ fusion subprocesses in the initial state collinear phase
space region are obtained as
\begin{eqnarray}
\label{initial collinear cross section}
\sigma_{HC}^{qq}&=&\int\hat{\sigma}_{\sss LO}^{qq}
\left[\frac{\alpha_s}{2 \pi} \frac{\Gamma(1-\epsilon)}{\Gamma(1-2
\epsilon)}\left(\frac{4 \pi
\mu_r^2}{\hat{s}}\right)^{\epsilon}\right]\left\{
\tilde{G}_{q/A}(x_A,\mu_f)G_{\bar q/B}(x_B,\mu_f)+G_{\bar
q/A}(x_A,\mu_f)\tilde{G}_{q/B}(x_B,\mu_f)
\right.\nb \\
&+& \left. \sum_{\alpha=u,\bar u}^{d,\bar
d}\left[\frac{A_1^{sc}(\alpha \to \alpha
g)}{\epsilon}+A_0^{sc}(\alpha \to \alpha
g)\right]G_{q/A}(x_A,\mu_f)G_{\bar q/B}(x_B,\mu_f)+(A
\leftrightarrow B)\right\}dx_Adx_B,\\
\sigma_{\sss HC}^{gg}&=&\int\hat{\sigma}_{\sss LO}^{gg}
\left[\frac{\alpha_s}{2 \pi} \frac{\Gamma(1-\epsilon)}{\Gamma(1-2
\epsilon)}\left(\frac{4 \pi
\mu_r^2}{\hat{s}}\right)^{\epsilon}\right]\frac{1}{2}\left\{
\tilde{G}_{g/A}(x_A,\mu_f)G_{g/B}(x_B,\mu_f)+G_{g/A}(x_A,\mu_f)\tilde{G}_{g/B}(x_B,\mu_f)
\right.\nb \\
&+& \left. \left[\frac{A_1^{sc}(g\to gg)}{\epsilon}+A_0^{sc}(g\to
gg)\right]G_{g/A}(x_A,\mu_f)G_{g/B}(x_B,\mu_f)+(A \leftrightarrow
B)\right\}dx_Adx_B,
\end{eqnarray}
where
\begin{eqnarray}
A_0^{sc}&=&A_1^{sc} \ln\left(\frac{\hat{s}}{\mu_f^2}\right), \nb \\
\tilde{G}_{\alpha/A,B}(x,\mu_f)&=&\int^{1-\delta_s}_x
\frac{dy}{y}G_{\alpha/A,B}(x/y,\mu_f)\tilde{P}_{\alpha \alpha}(y),
~~~~(\alpha=u,\bar u,d,\bar d,g),
\end{eqnarray}
with
\begin{eqnarray}
\tilde{P}_{\alpha \alpha}(y)=P_{\alpha \alpha}(y)
\ln\left(\frac{\delta_c}{2}\frac{(1-y)^2}{y}\frac{\hat{s}}{\mu_f^2}\right)-P'_{\alpha
\alpha}(y),~~~~(\alpha=u,\bar u,d,\bar d,g).
\end{eqnarray}

\par
{\bf 2. Hard Light-quark Emission Subprocesses $(q,\bar q)g \to
t\bar b H^-+(q,\bar q)$}

\par
The method in the calculation of the hard light-quark emission
subprocesses $(q,\bar q)g\rightarrow t\bar b H^-+(q,\bar q)$ is
similar to that for hard gluon emission subprocesses. In the
collinear region, the initial state parton $i(i=u,d,\bar u,\bar
d,g)$ is considered to split into a hard parton $i'$ and a
collinear light-quark, $i \to i'q$, with $p_{i'}=zp_i$ and
$p_6=(1-z)p_i$. Let the hard light-quark be emitted collinear to
one of the incoming partons, the collinear region is then defined
as:
\begin{eqnarray}
\frac{2p_1\cdot p_6}{E_6\sqrt{\hat{s}}} < \delta_c~~~or~~~~
\frac{2p_2\cdot p_6}{E_6\sqrt{\hat{s}}} < \delta_c.
\end{eqnarray}
The collinear singularity of $\hat\sigma^{qg}_{real}$ can be
written as:
\begin{eqnarray}
\label{initial collinear} \hat\sigma^{qg}_{\sss HC}&=&
\left[\frac{\alpha_s}{2 \pi} \frac{\Gamma(1-\epsilon)}{\Gamma(1-2
\epsilon)}\left(\frac{4 \pi \mu_r^2}{s'}\right)^{\epsilon}\right]
\left(-\frac{1}{\epsilon}\right)\delta_c^{-\epsilon}
\left\{ \int_0^{1-\delta_s}dz\left[\frac{(1-z)^2}{2z}\right]^{-\epsilon} \right.\nb \\
&&\left. \times P_{qg}(z,\epsilon)\hat\sigma_{LO}^{gg}(gg\to t\bar
b H^-)+P_{gq}(z,\epsilon)\hat\sigma_{LO}^{qq}(q\bar q\to t\bar b
H^-)\right\}.
\end{eqnarray}
with
\begin{eqnarray}
P_{ii'}(z,\epsilon)&=&P_{ii'}(z)+ \epsilon P'_{ii'}(z), \nb \\
P_{gq}(z)&=&\frac{1}{2}\left[z^2+(1-z)^2\right],~~~~~~~~~~~~~~~
P'_{gq}(z)=-z(1-z), \nb \\
P_{qg}(z)&=&\frac{N^2-1}{2N}\left(\frac{1+(1-z)^2}{z}\right),~~~~~~~~
P'_{qq}(z)=-\frac{N^2-1}{2N}z.
\end{eqnarray}
Using the $\overline{MS}$ scheme, the scale dependent distribution
function can be written as:
\begin{eqnarray}
G_{i'/A}(x,\mu_f)&=& G_{i'/A}(x)
+\left(-\frac{1}{\epsilon}\right)\left[\frac{\alpha_s}{2 \pi}
\frac{\Gamma(1-\epsilon)}{\Gamma(1-2 \epsilon)}\left(\frac{4\pi
\mu_r^2}{\mu_f^2}\right)^{\epsilon}\right]\int_x^{1}\frac{dz}{z}P_{ii'}(z)G_{i/A}(x/z).
\end{eqnarray}
And we can get the expression for the initial state collinear
contribution at $O(\alpha_s)^3$ order:
\begin{eqnarray}
\sigma^{qg}_{\sss HC} &=& \frac{\alpha_s}{2\pi}
\sum_{i=q,\bar{q}}\int dx_Adx_B \left\{ \int_{x_A}^{1}\frac{dz}{z}
G_{i/A}(\frac{x_A}{z},\mu) G_{g/B}(x_B,\mu)
\times \right. \nonumber \\
&&\left. \hat{\sigma}_{LO}^{gg}(x_A, x_B,\mu) \left[P_{ig}(z)
\ln\left(\frac{s}{\mu^2}\frac{(1-z)^2}{z}\frac{\delta_c}{2}\right)-
P^{\prime}_{ig}(z) \right]\right.\nonumber\\
&+& \left. \int_{x_A}^{1}\frac{dz}{z} G_{g/A}(\frac{x_A}{z},\mu)
G_{i/B}(x_B,\mu)
\times \right. \nonumber\\
&&\left. \hat{\sigma}_{LO}^{q\bar{q}}(x_A, x_B,\mu) \left[
{P}_{gi}(z) \ln\left(\frac{s}{\mu^2}\frac{(1-z)^2}{z}
\frac{\delta_c}{2}\right)- {P}^{\prime}_{gi}(z)\right]
+(A\leftrightarrow B)\right\}.
\end{eqnarray}

\par
\subsection{Total NLO Cross Section}
\par
The final result for the $O(\alpha_s)$ NLO QCD corrected cross
section part of $qq$ annihilation subprocesses can be written as:
\begin{eqnarray}
\sigma_{\sss NLO}^{qq}&=&\int dx_Adx_BG_{q/A}(x_A,\mu) G_{\bar
q/B}(x_B,\mu)\left[\hat\sigma_{\sss
LO}^{qq}(x_A,x_B,\mu)+\hat\sigma_{virtual}^{qq}(x_A,x_B,\mu)
+\hat\sigma_{soft}^{qq}(x_A,x_B,\mu) \right. \nonumber\\
&+&\left. (A\leftrightarrow B)\right]+\sigma_{\sss HC}^{qq}+\int
dx_Adx_B\left[G_{q/A}(x_A,\mu) G_{\bar
q/B}(x_B,\mu)\hat\sigma_{\overline {\sss
HC}}^{qq}(x_A,x_B,\mu)+(A\leftrightarrow B)\right].
\end{eqnarray}
And the NLO QCD corrected cross section part for $gg$ fusion
subprocess has the expression as:
\begin{eqnarray}
\sigma_{NLO}^{gg}&=&\frac{1}{2}\int dx_Adx_BG_{g/A}(x_A,\mu)
G_{g/B}(x_B,\mu)[\hat\sigma_{LO}^{gg}(x_A,x_B,\mu)+\hat\sigma_{virtual}^{gg}(x_A,x_B,\mu)
+\hat\sigma_{soft}^{gg}(x_A,x_B,\mu) \nonumber\\
&+&(A\leftrightarrow B)]+\sigma_{HC}^{gg}+\frac{1}{2}\int
dx_Adx_B\left[G_{g/A}(x_A,\mu)
G_{g/B}(x_B,\mu)\hat\sigma_{\overline {\sss
HC}}^{gg}(x_A,x_B,\mu)+(A\leftrightarrow B)\right].
\end{eqnarray}
The cross section of $(q,\bar q)g \to t\bar b H^-+(q,\bar q)$
($q=u,d$) can be written as:
\begin{eqnarray}
\sigma_{NLO}^{qg}&=&\sigma^{qg}_{HC}+\sum_{i=q,\bar{q}}\int
dx_Adx_B[G_{i/A}(x_A,\mu) G_{g/B}(x_B,\mu)\hat\sigma_{\overline
{HC}}^{qg}(x_A,x_B,\mu)+(A \leftrightarrow B)].
\end{eqnarray}
with the hard-non-collinear partonic cross section given by
\begin{eqnarray}
\hat{\sigma}_{\overline {\sss HC}}^{ij}=\int_{\overline{\rm
HC}}\overline{\sum}|M(ij \to t\bar b H^-g(q,\bar q))|^2 d \Phi_4.
\end{eqnarray}
where $d\Phi_4$ denotes the four-particle phase space element.
\par
Finally, the NLO QCD corrected total cross sections for
$pp/p\bar{p} \to t\bar b H^-+X$ can be obtain by using the
formula:
\begin{eqnarray}
\sigma_{\sss
NLO}=\sigma_{NLO}^{qq}+\sigma_{NLO}^{gg}+\sigma_{NLO}^{qg}.
\end{eqnarray}
In our calculation, we have checked the cancellations of the UV
and IR divergence analytically, and the final results are both UV-
and IR-finite.

\par
\section{ Numerical Results and Discussions}

\par
In our numerical calculation, we adopt the CTEQ6M\cite{CTEQ}
parton distribution functions and the 2-loop evolution of
$\alpha_s(\mu)$ to evaluate the hadronic NLO QCD corrected cross
sections with $\alpha_s^{NLO}(M_Z)=0.118$, while for the hadronic
LO cross sections we use the CTEQ6L1 parton distribution functions
and the one-loop evolution of $\alpha_s(\mu)$. We take the SM
parameters as: $\alpha_{{\rm ew}}(M_Z)^{-1} = 127.918$, $m_W =
80.423~GeV$, $m_Z = 91.18~GeV$, $m_t = 175~GeV$, $m_b = 4.62~GeV$.
As a numerical demonstration, in this work we refer to the
Snowmass point SPS1b for the relevant MSSM
parameters\cite{snowmass}, if there is no other statement. The
MSSM parameters in this benchmark are given by:
\begin{eqnarray}
\tan\beta = 30,~~~\mu= 495.6~GeV,~~~A_t = -729.3~GeV, \nonumber\\
A_b=-987.4~GeV,~~~m_{\tilde g} = 916.1~GeV,~~~m_{\tilde q_L} =
762.5~GeV, \nonumber\\
m_{\tilde b_R} =780.3~GeV,~~~m_{\tilde t_R} = 670.7~GeV.
\end{eqnarray}
The charged Higgs boson mass is taken as free parameter. The
calculations are carrying out at the upgraded Tevatron with the
$p\bar p$ colliding energy $\sqrt{s}=2~TeV$ and the LHC with $pp$
colliding energy $\sqrt{s}=14~TeV$.

\par
As the check of the correctness of our numerical calculation, we
plot Fig.4 to show the dependence of $\sigma_{real}^{gg}$ on
$\delta_c$ and $\delta_s$ at the LHC. Fig.4(a) shows the
$\sigma_{real}^{gg}$ as the functions of $\delta_c$ by taking
$\delta_s=10^{-4}$, $m_{H^-}=310~GeV$ and $\delta_c$ varying from
$10^{-6}$ to $10^{-4}$. Fig.4(b) presents the curves of
$\sigma_{real}^{gg}$ versus $\delta_s$ with $\delta_c=10^{-5}$,
$m_{H^-}=310~GeV$ and $\delta_s$ running from $10^{-5}$ to
$10^{-3}$. Both figures show that our results of
$\sigma_{real}^{gg}(gg \to t\bar bH^-+g)$ is independent of cutoff
$\delta_c$ and $\delta_s$. Actually, our calculation proves also
that the real gluon emission cross section for $q\bar q$
annihilation channel is independent on cutoff $\delta_c$ and
$\delta_s$, and the cross sections for $qg$ and $\bar q g$ fusion
channels(as shown in (\ref{subprocesses4})) are independent on
$\delta_c$ too.

\par
In following calculation, we require the final anti-bottom quark
to have a transverse momentum($p_T^b$) being larger than $20~GeV$
and a pseudorapidity $|\eta_b| \leq 2$ for the Tevatron and less
than $|\eta_b| \leq 2.5$ for the LHC, unless other statement is
given. We present the dependence of the cross section on the
renormalization/factorization scale $Q/Q_0$ (where we take
$Q\equiv\mu_r=\mu_f$ for simplicity and define
$Q_0=(m_t+m_{H^-}+m_b)/2$ ) in Fig.5(a1-a2) and Fig.5(b1-b2)
taking $m_{H^-}=175~GeV$ at the Tevatron and $m_{H^-}=250~GeV$ at
the LHC, separately. In Fig.5(a1) and Fig.5(b1) we depict the
curves for total cross sections at leading-order $\sigma_{LO}$ and
$\sigma_{NLO}$ including NLO corrections for the processes $p\bar
p/pp \to t\bar bH^-+X$ at the Tevatron and the LHC. They show that
the $\sigma_{NLO}$ in the MSSM is less dependent on the
normalization/factorization scale $Q/Q_0$ than $\sigma_{LO}$ both
at the Tevatron and the LHC, especially in Fig.5(b1) the curve for
$\sigma_{NLO}$ at the LHC is nearly independent of $Q/Q_0$. We can
see from Fig.5(a1) that the NLO QCD corrections suppress the LO
cross section of the process $p\bar p \to t\bar bH^-+X$ at the
Tevatron. Fig.5(a2) shows that at the Tevatron the dominant
contribution to $\sigma_{NLO}$ is from the subprocess $q\bar{q}
\to t\bar bH^-$, while Fig.5(b2) demonstrates that the main
contributions are coming from subprocess $gg \to t\bar{b} H^-$ at
the LHC. From Fig.5(a1) we can see that if $Q$ goes down to a very
low value, i.e., $Q<<Q_0$, the curve for $\sigma_{NLO}$ tends to
have a negative value. The reason is that large logarithmic
corrections spoil the convergence of perturbation theory in the
proton-antiproton colliding energy at the Tevatron. From
Fig.6(a1-a2) and (b1-b2), we conclude that the dependence of the
NLO QCD corrected cross section $\sigma_{NLO}$ on the scale $Q$ is
significantly reduced comparing with $\sigma_{LO}$, especially at
the LHC, there the $\sigma_{NLO}$ is very stable in a large range
of $Q$.

\par
In the following figures, we fix the value of the
renormalization/factorization scale being $Q=Q_0=\mu_r=\mu_f$. In
Fig.6(a1) and (b1) we depict the LO and total NLO QCD corrected
cross sections, $\sigma_{LO}$ and $\sigma_{NLO}$, in the MSSM as
the functions of $m_{H^-}$ at the Tevatron and the LHC,
respectively. The corresponding relative NLO QCD corrections
$\delta(\equiv\frac{\sigma_{NLO}-\sigma_{LO}}{\sigma_{LO}})$
versus $m_{H^-}$ are plotted in Fig.6(a2) and (b2), separately.
From these figures, we can see that the cross sections
$\sigma_{NLO}$ and $\sigma_{LO}$ decrease rapidly as $m_{H^-}$
varies in the range from $175GeV$ to $300GeV$ at the Tevatron and
from $175GeV$ to $550GeV$ at the LHC. We can read from Fig.6(a1)
that when $m_{H^-}$ increases from $175~GeV$ to $300~GeV$, the
total NLO QCD corrected cross section $\sigma_{NLO}$ in the MSSM
decreases from $0.66~fb$ to $0.02~fb$ at the Tevatron. From
Fig.6(a2) we can read when $m_{H^-}$ increases from $175~GeV$ to
$550~GeV$, $\sigma_{NLO}$ decreases from $160~fb$ to $3~fb$ at the
LHC. The absolute value of the relative correction $\delta$ in
Fig.6(a2) decreases slightly form $39\%$ to $32\%$ at the
Tevatron. While the absolute value of the relative correction
$\delta$ in Fig.6(b2) increases rapidly with the increment of the
charged Higgs boson mass at the LHC, it reaches the value of
$44\%$ when $m_{H^-}=550~GeV$.

\par
In Fig.7(a-b), we present the total LO and NLO QCD corrected cross
sections $\sigma_{LO}$ and $\sigma_{NLO}$ in the MSSM as the
functions of the transverse momentum of anti-bottom quark cut
$p_{T,cut}^b$, taking $m_{H^-}=175~GeV$ at the Tevatron and
$m_{H^-}=250~GeV$ at the LHC, respectively. In these two figures,
we can see that the cross sections $\sigma_{NLO}$ and
$\sigma_{LO}$ decrease as $p_{T,cut}^b$ goes up from $10~GeV$ to
$25~GeV$. We can read from Fig.7(a-b) that when $p_{T,cut}^b$
increases from $10~GeV$ to $25~GeV$, the total NLO QCD corrected
cross section $\sigma_{NLO}$ in the MSSM decreases roughly from
$0.87~fb$($55~fb$) to $0.56~fb$($37~fb$) for the Tevatron(LHC).

\par
Fig.8(a1) and Fig.8(b1) show the LO and total NLO QCD corrected
cross sections $\sigma_{LO}$ and $\sigma_{NLO}$ in the MSSM as the
functions of $\tan\beta$ with $m_{H^-}=175~GeV$ at the Tevatron
and $m_{H^-}=250~GeV$ at the LHC, respectively. When $\tan\beta$
has a small value, the NLO QCD corrected cross section decreases
from $0.45~fb$($30~fb$) to $0.15~fb$($127~fb$) at the
Tevatron(LHC), as $\tan\beta$ goes up from $4$ to $12$. The NLO
QCD corrected cross sections $\sigma_{NLO}$ increases from
$0.15~fb$($12~fb$) to $2~fb$ ($132~fb$) for the Tevatron(LHC), as
$\tan\beta$ varies from $12$ to $50$. The corresponding relative
corrections $\delta$ are plotted in Fig.8(a2) and Fig.8(b2). The
$\delta$ can be beyond $-50\%$ at the Tevatron, and approach
$-40\%$ at the LHC at the position of $\tan\beta \sim 15$.

\par
In Fig.9, we depict the distributions of the transverse momenta of
the final states($p_T^b$, $p_T^t$ and $p_T^{H^-}$) with
$m_{H^-}=250~GeV$ at the LHC and $m_{H^-}=175~GeV$ at the
Tevatron, separately. In Fig.9(a1-a2, b1-b2 ,c1-c2), we show the
distributions of the differential cross sections
$d\sigma_{LO,NLO}/dp_T^b$, $d\sigma_{LO,NLO}/dp_T^t$ and
$d\sigma_{LO,NLO}/dp_T^{H^-}$ at the LHC and the Tevatron,
respectively. These figures demonstrate that the NLO QCD
corrections significantly modify the leading-order distributions
of the differential cross sections $d\sigma_{LO}/dp_T^b$,
$d\sigma_{LO}/dp_T^t$ and $d\sigma_{LO}/dp_T^{H^-}$ at hadron
colliders. We find that in the low $p_T^b$ region the differential
cross sections $d\sigma_{LO,NLO}/dp_T^b$ can be very large,
especially at the LHC.

\par
\section{ Summary}

\par
In this paper we calculated the NLO QCD corrections to the
processes $p\bar{p}/pp \to t\bar{b} H^-+X$ in the MSSM at the
Tevatron and the LHC. We investigated the contributions of the NLO
QCD corrections to the total cross sections and the distributions
of the transverse momenta of final particles(anti-bottom-quark,
t-quark and charged Higgs-boson), and found the NLO QCD
corrections significantly modify the corresponding LO differential
cross sections. We analyzed the dependence of the NLO QCD
corrected cross sections and the corresponding relative
corrections on the renormalization/factorization scale $Q$,
charged Higgs-boson mass $m_{H^-}$, transverse momentum cut of
anti-bottom quark $p_{T,cut}^b$, and $\tan\beta$, respectively.
Our numerical results show that the theoretical NLO QCD
corrections in the MSSM reduce the dependence of the cross section
on the factorization and normalization scales, especially the NLO
QCD corrected total cross sections at the LHC are nearly
independent of these scales, and the relative correction is
obviously related to $m_{H^-}$ and $\tan\beta$ at both the
Tevatron and the LHC. We find the total NLO QCD relative
corrections $\delta$ can be beyond $-50\%$ at the Tevatron and
approach $-40\%$ at the LHC.

\paragraph{Acknowledgments}
This work was supported in part by the National Natural Science
Foundation of China and special fund sponsored by China Academy of
Science.

\begin{figure}[htp]
\centering
\includegraphics[width=3.2in,height=3in]{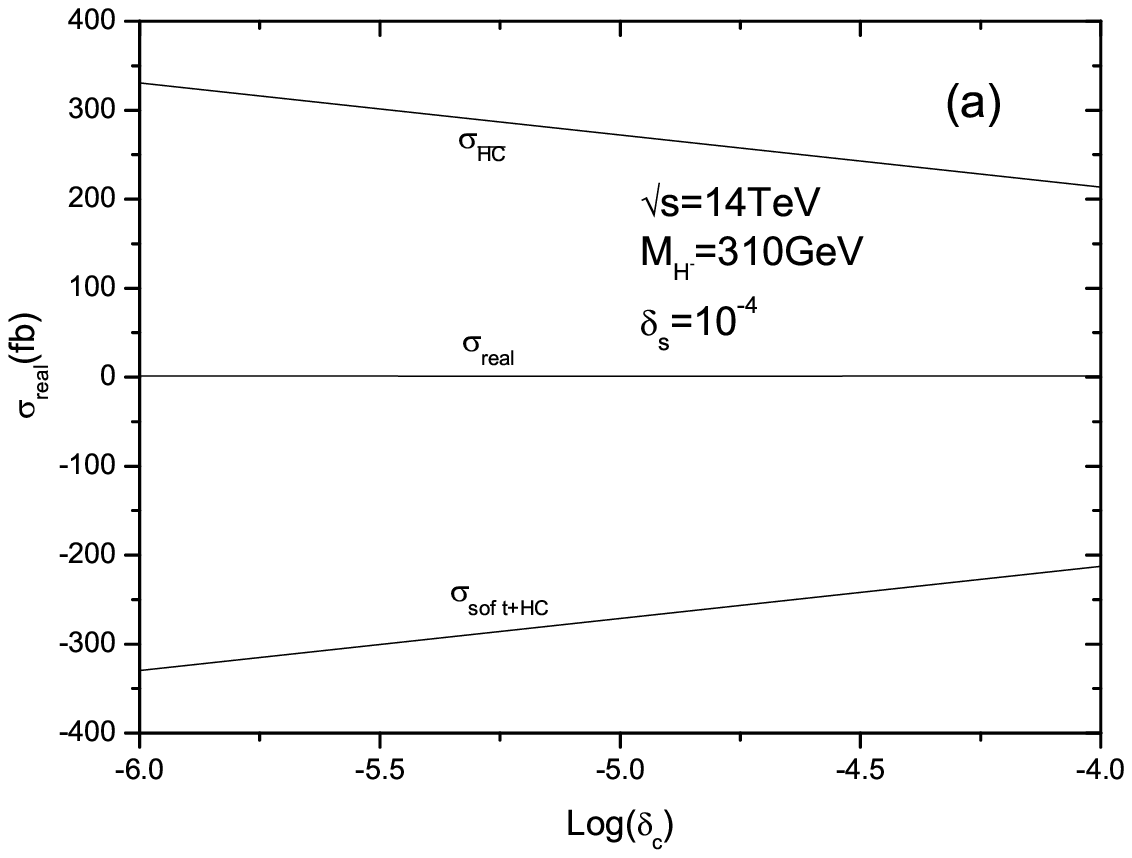}%
\hspace{0in}%
\includegraphics[width=3.2in,height=3in]{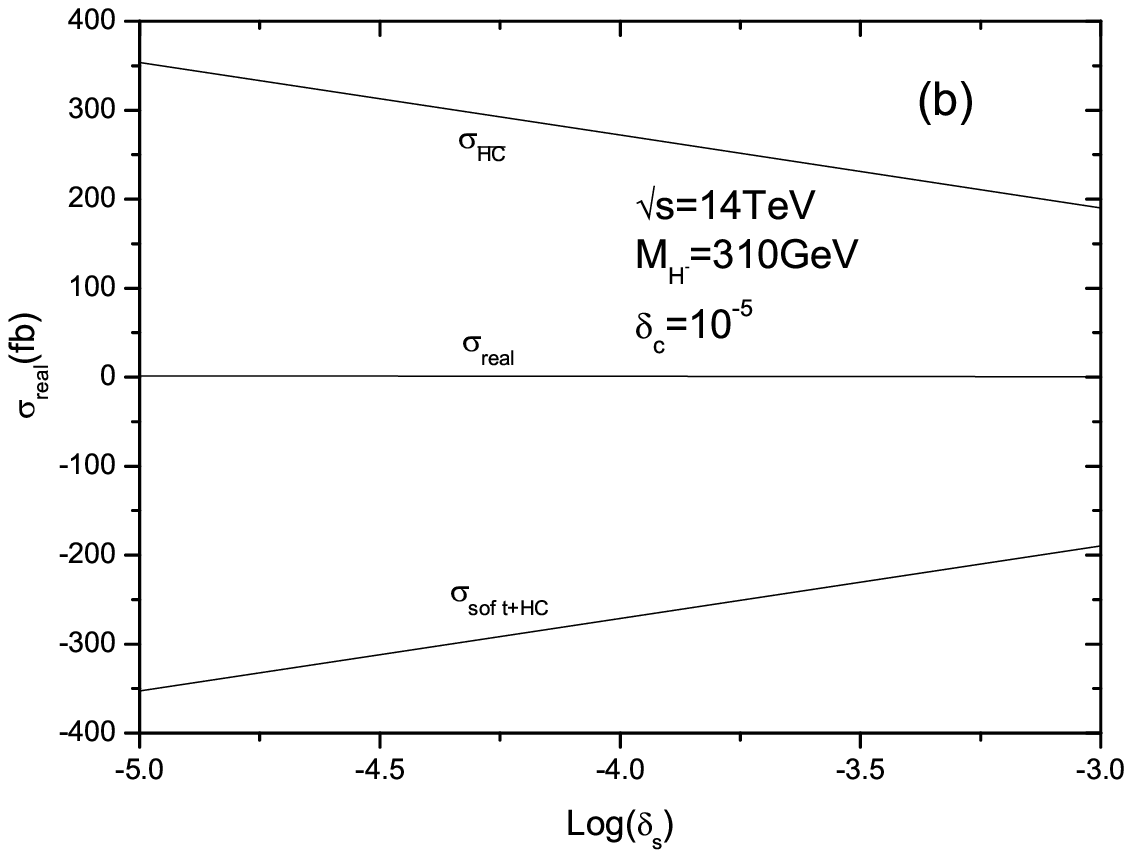}
\caption{The cutoff dependence of $\sigma_{real}^{gg}$ with
$M_{H^-}=310~GeV$ at the Snowmass point SPS1b at the LHC. Fig.4(a)
shows the $\delta_c$ dependence with $\delta_s=10^{-4}$, and
Fig.4(b) shows the $\delta_s$ dependence with $\delta_c=10^{-5}$.}
\end{figure}

\begin{figure}[htp]
\centering
\includegraphics[width=3.2in,height=3in]{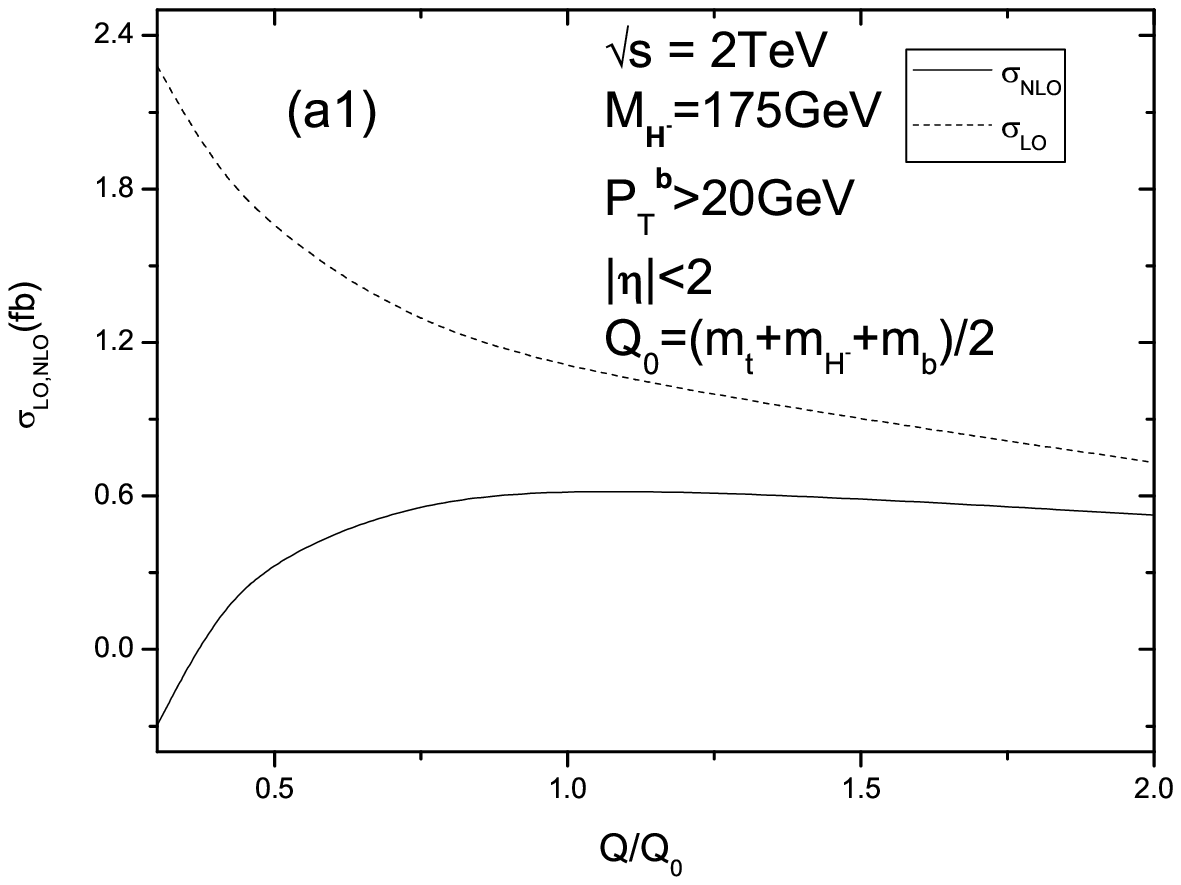}%
\hspace{0in}%
\includegraphics[width=3.2in,height=3in]{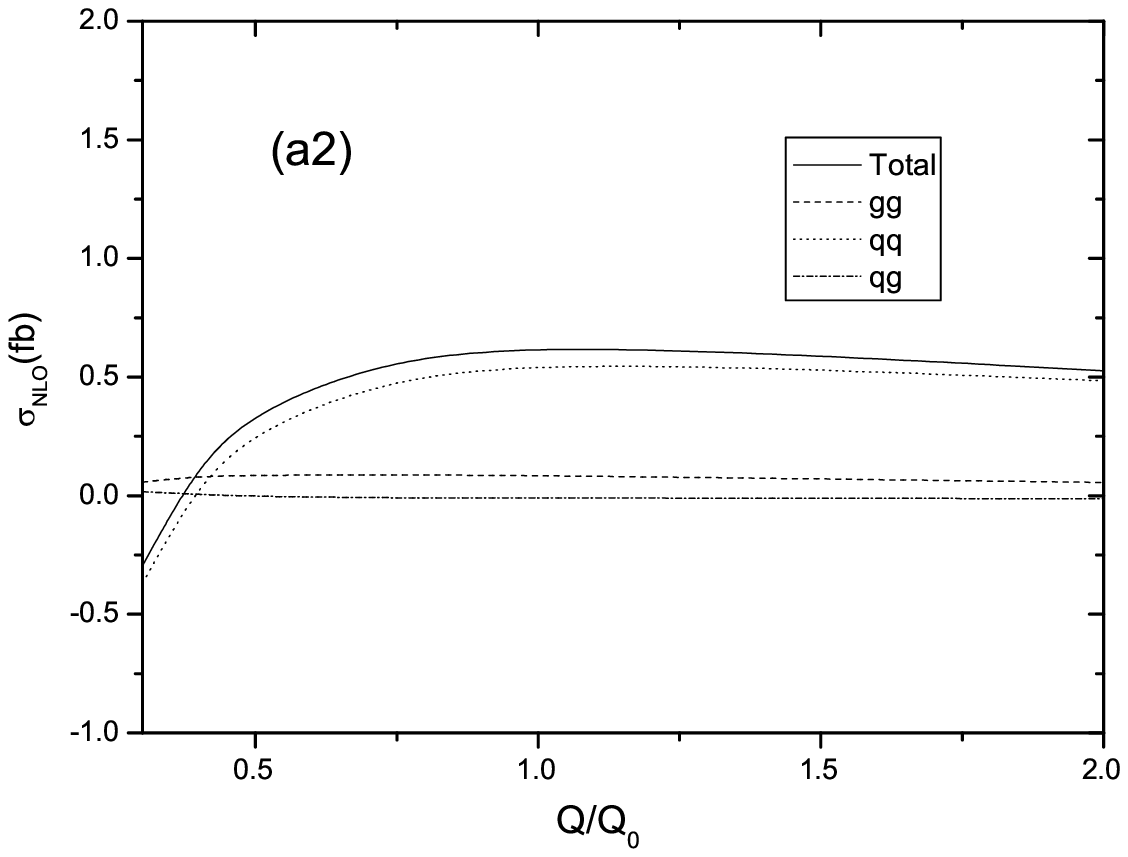}
\hspace{0in}%
\includegraphics[width=3.2in,height=3in]{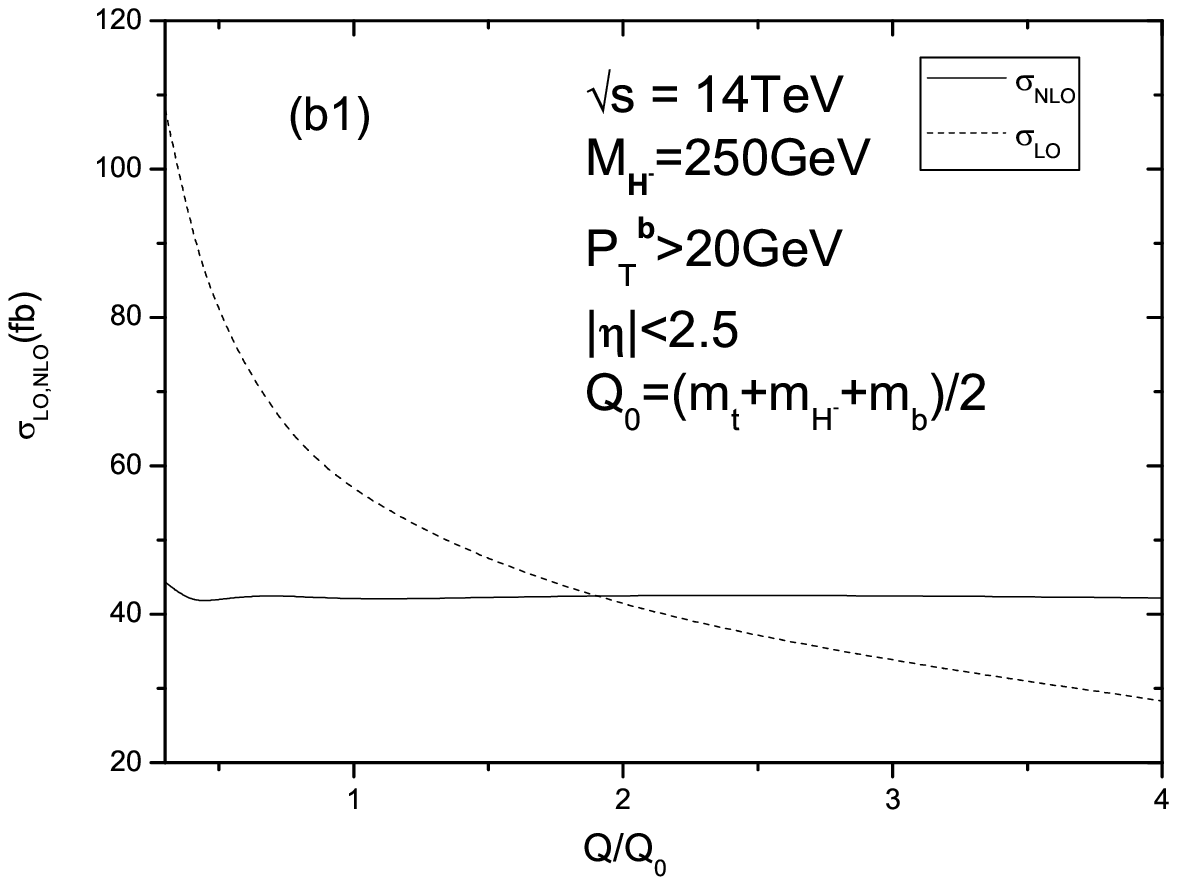}%
\hspace{0in}%
\includegraphics[width=3.2in,height=3in]{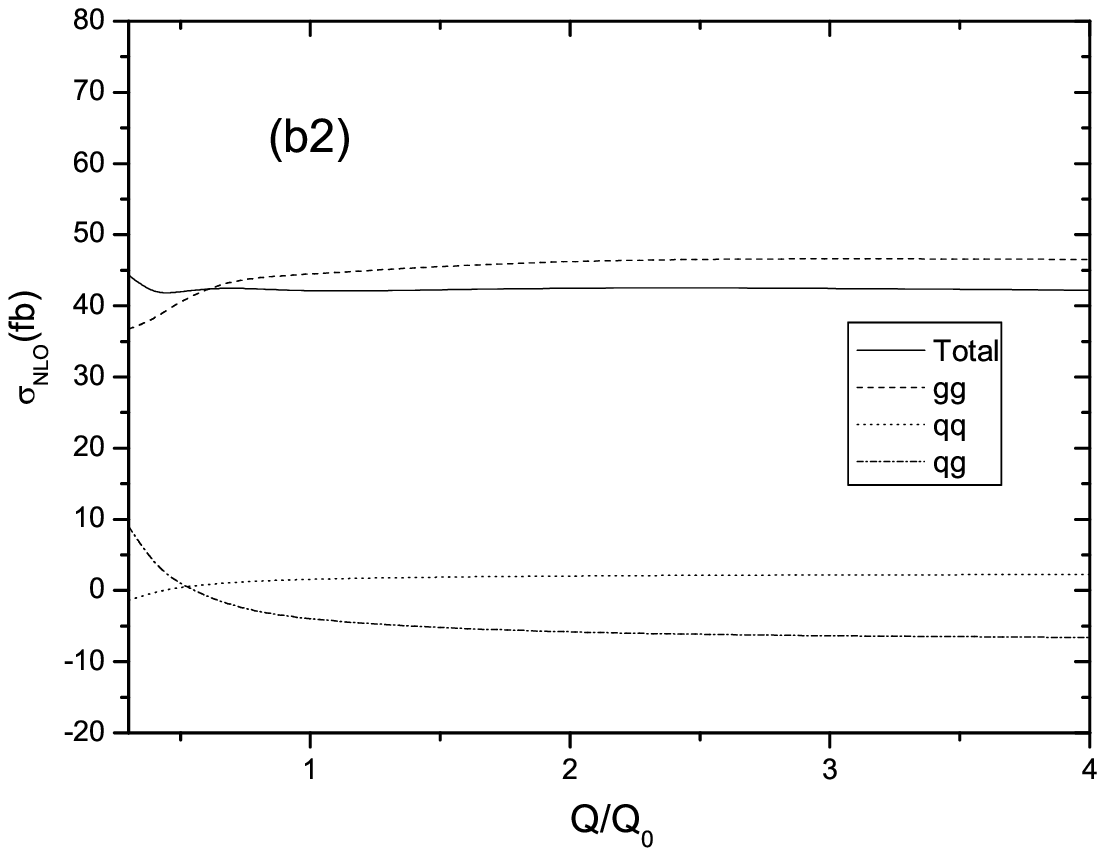}
\caption{The total cross sections $\sigma_{LO}$ and $\sigma_{NLO}$
for the processes $pp/p\bar p \to t\bar bH^-+X$ as the functions
of the renormalization/factorization scale $Q$ with
$m_{H^-}=175~GeV$, $|\eta_b| \leq 2$ at the Tevatron and
$m_{H^-}=250~GeV$, $|\eta_b| \leq 2.5$ at the LHC, are shown in
Fig.5(a1) and Fig.5(b1) respectively. The contribution parts to
the total cross sections $\sigma_{NLO}(pp/p\bar p \to t\bar
bH^-+X)$ from the related $qq$, $gg$, $qg$ fusion subprocesses as
the functions of the renormalization/factorization scale $Q$ with
$m_{H^-}=175~GeV$ at the Tevatron and $m_{H^-}=250~GeV$ at the
LHC, are shown in Fig.5(a2) and Fig.5(b2) separately. }
\end{figure}

\begin{figure}[htp]
\centering
\includegraphics[width=3.2in,height=3in]{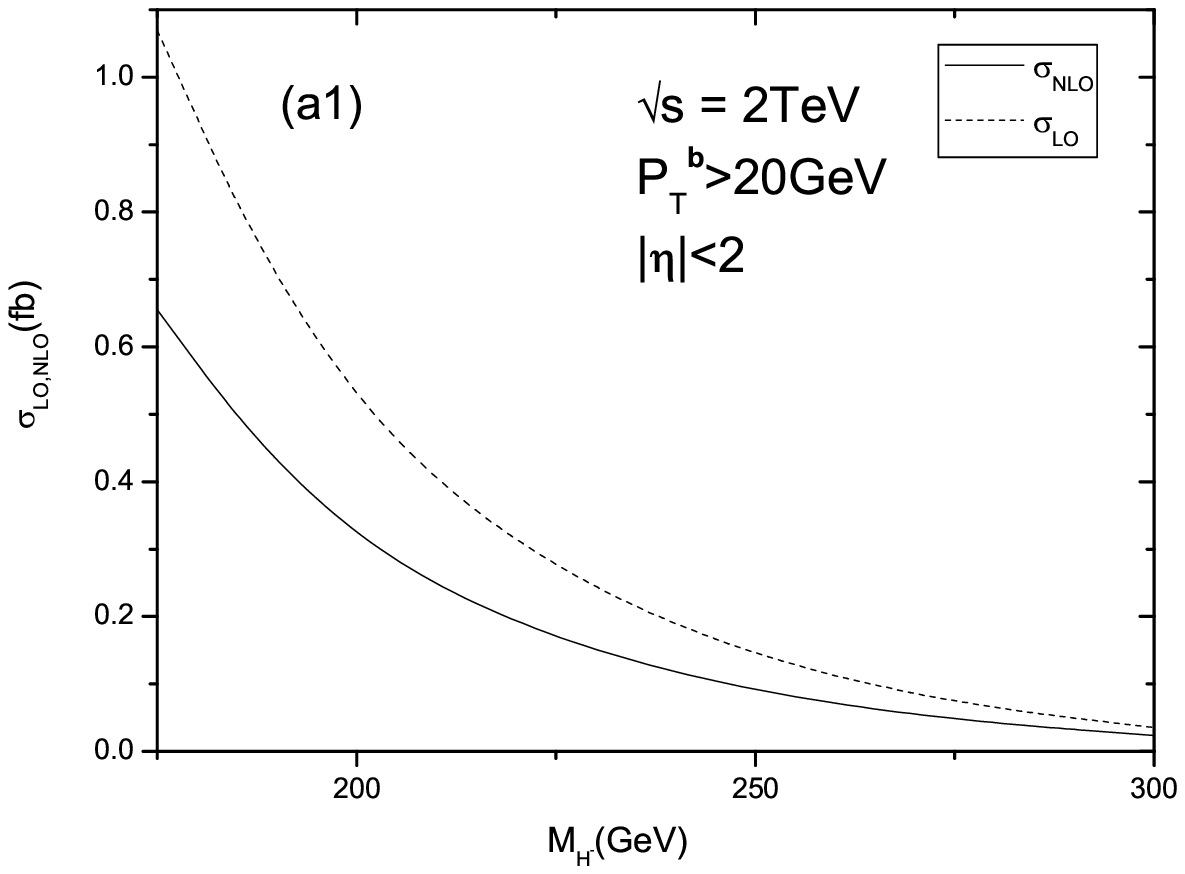}%
\hspace{0in}%
\includegraphics[width=3.2in,height=3in]{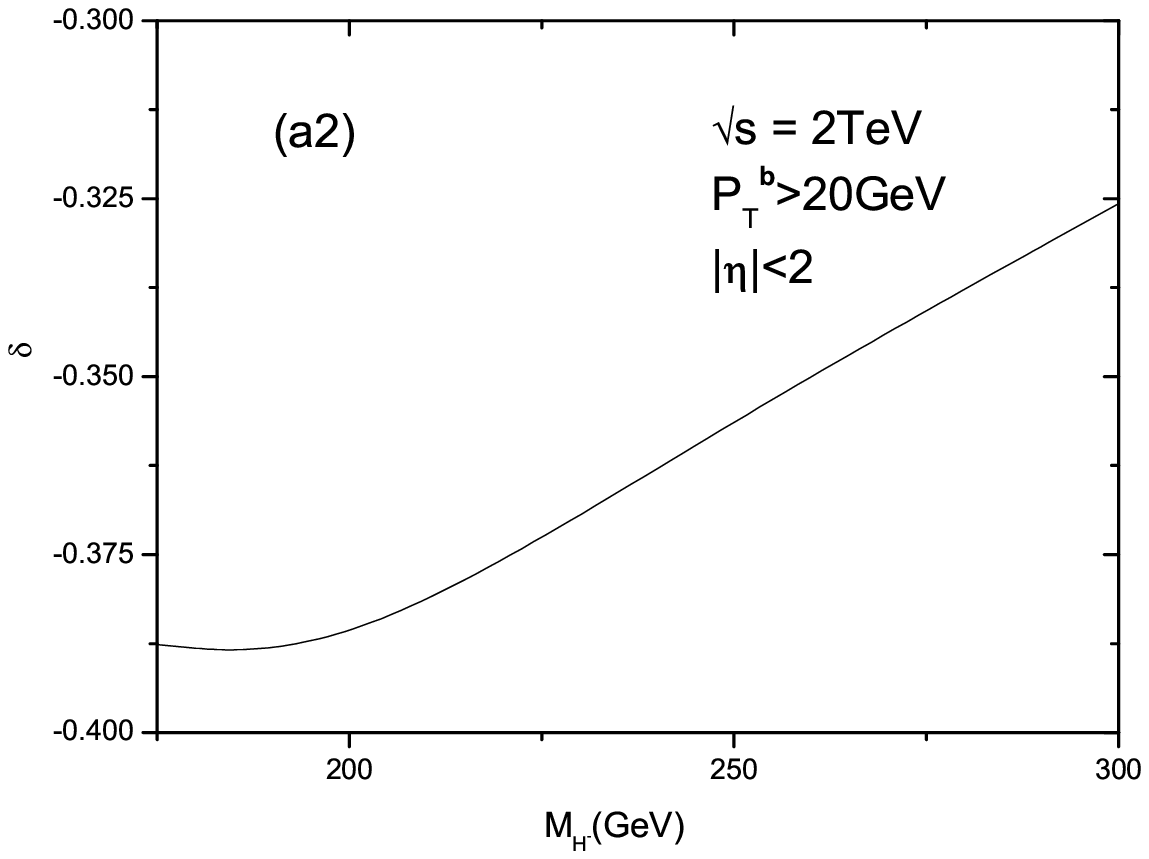}
\hspace{0in}%
\includegraphics[width=3.2in,height=3in]{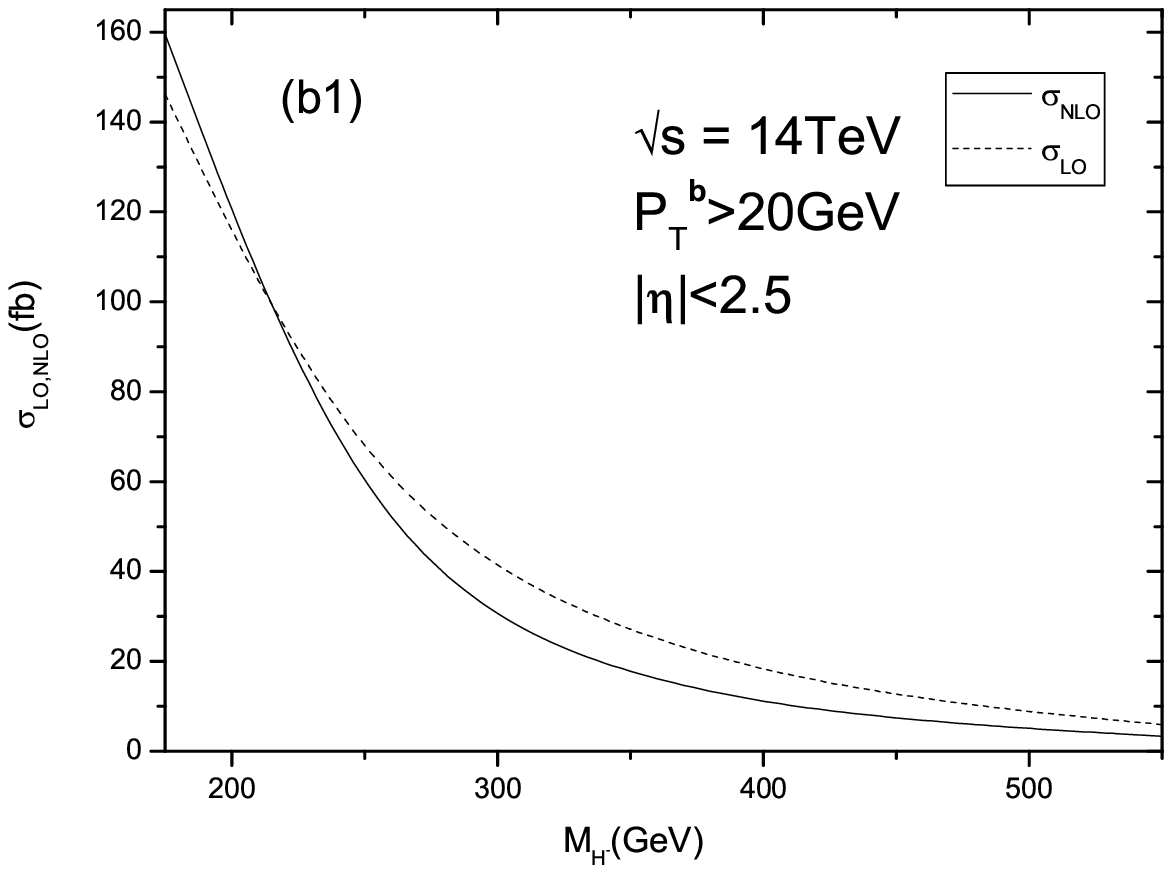}%
\hspace{0in}%
\includegraphics[width=3.2in,height=3in]{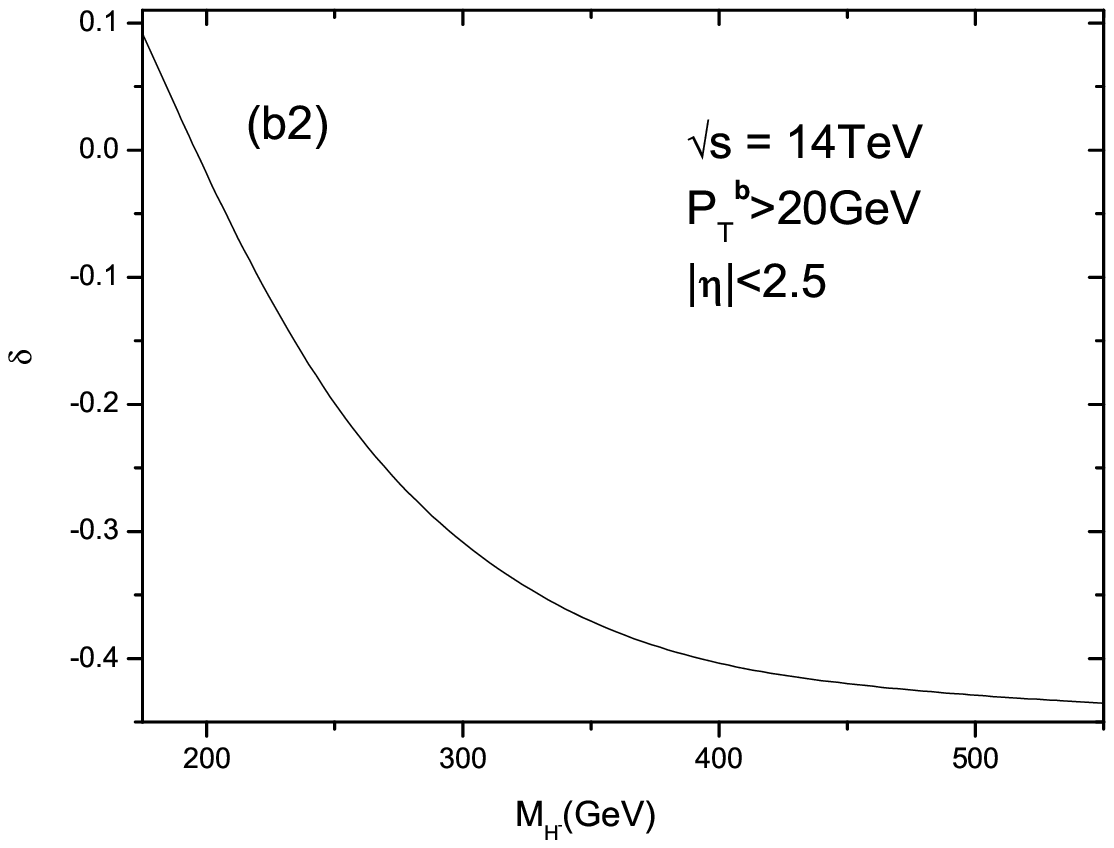}
\caption{The total NLO QCD corrected cross sections
($\sigma_{NLO}$) and the corresponding relative
corrections($\delta$) of the processes $p \bar{p}/pp \to t \bar
bH^-+X$ with $m_{H^-}=175~GeV$, $|\eta_b| \leq 2$ at the Tevatron
and $m_{H^-}=250~GeV$, $|\eta_b| \leq 2.5$ at the LHC, as the
functions of $m_{H^-}$. Fig.6(a1) and Fig.6(a2) are for the
process $p\bar p \to t\bar bH^-+X$ at the Tevatron and Fig.6(b1)
and Fig.6(b2) for the process $pp \to t\bar bH^-+X$ at the LHC.}
\end{figure}

\begin{figure}[htp]
\centering
\includegraphics[height=3in]{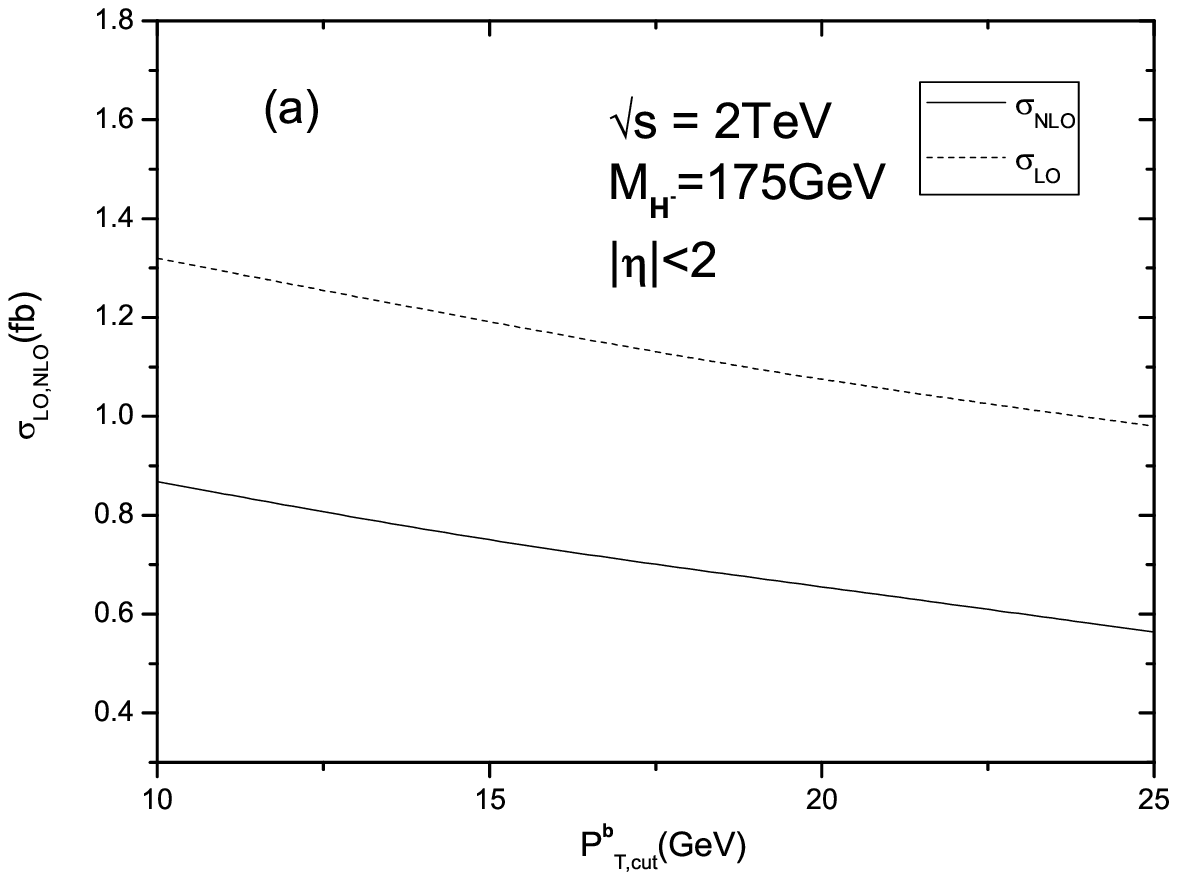}%
\hspace{0in}%
\includegraphics[height=3in]{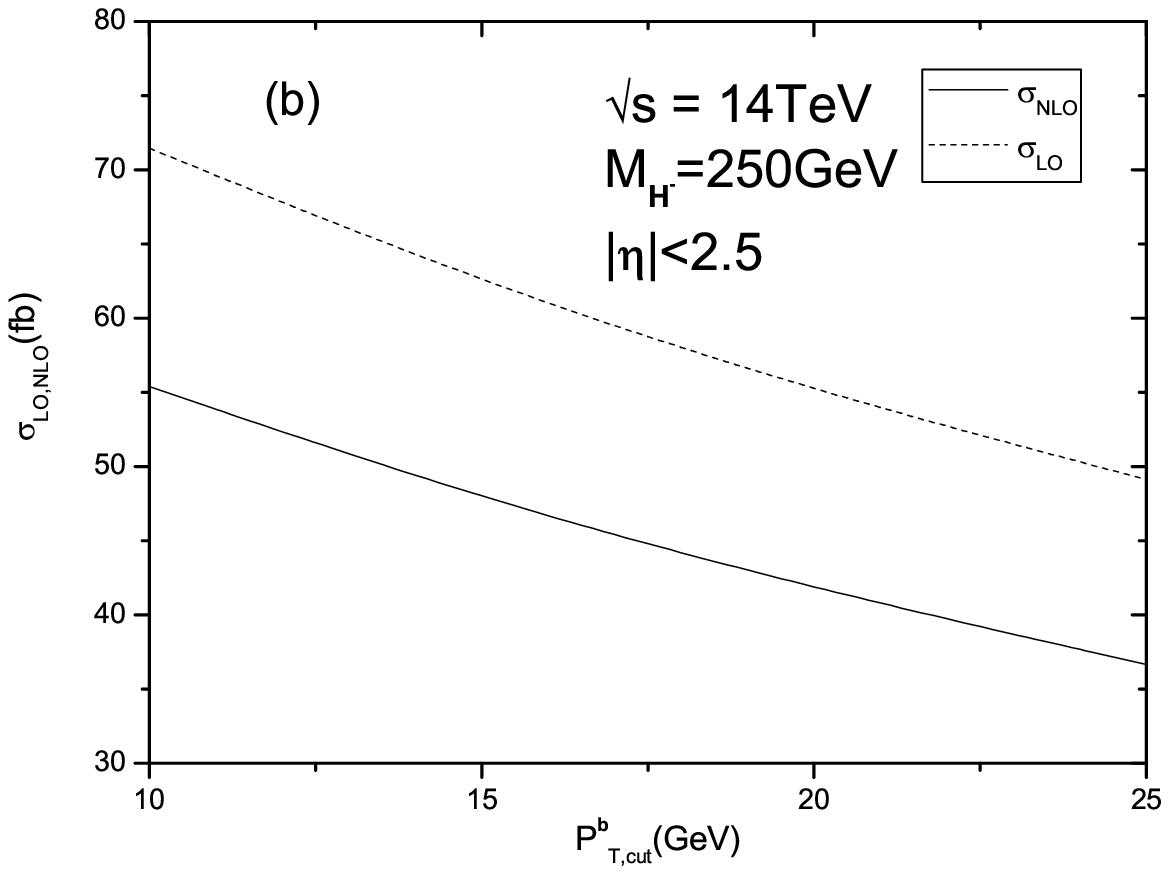}
\caption{The total NLO QCD corrected cross
sections($\sigma_{NLO}$) of the processes $p \bar{p}/pp \to t \bar
bH^-+X$ at the Tevatron and the LHC, as the functions of the
anti-bottom quark transverse momentum cut $p_{T,cut}^b$ with
$m_{H^-}=175~GeV$, $|\eta_b| \leq 2$ at the Tevatron and
$m_{H^-}=250~GeV$, $|\eta_b| \leq 2.5$ at the LHC. Fig.7(a) is for
the process $p\bar p \to t\bar bH^-+X$ at the Tevatron and
Fig.7(b) for the process $pp \to t\bar bH^-+X$ at the LHC.}
\end{figure}

\begin{figure}[htp]
\centering
\includegraphics[width=3.2in,height=3in]{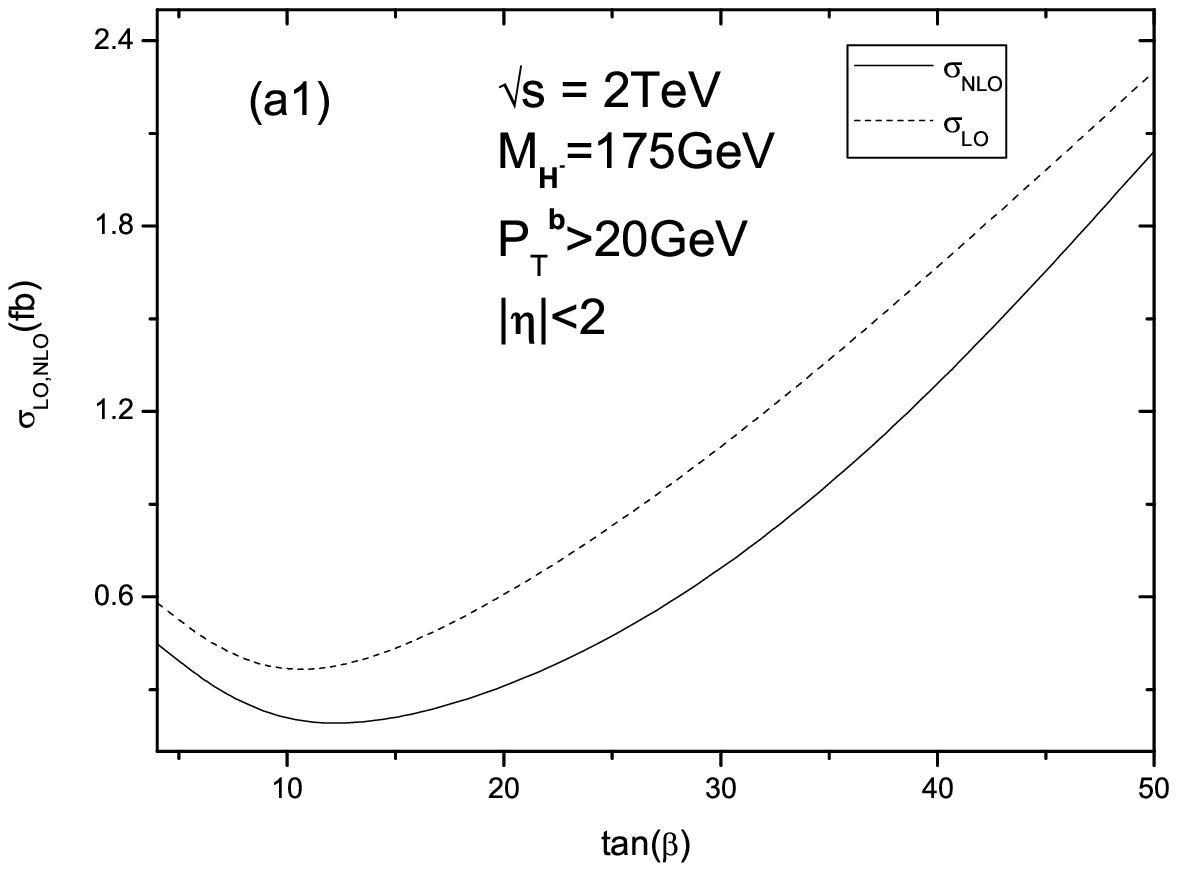}%
\hspace{0in}%
\includegraphics[width=3.2in,height=3in]{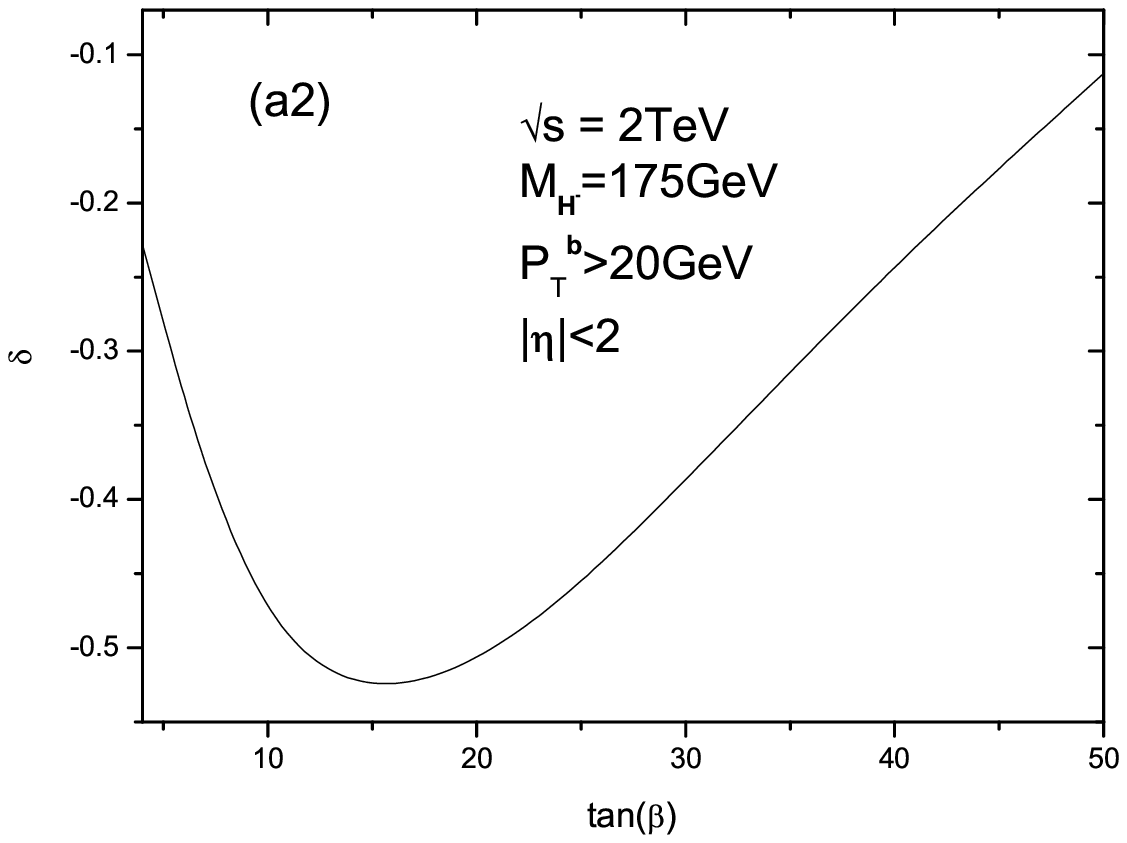}
\hspace{0in}%
\includegraphics[width=3.2in,height=3in]{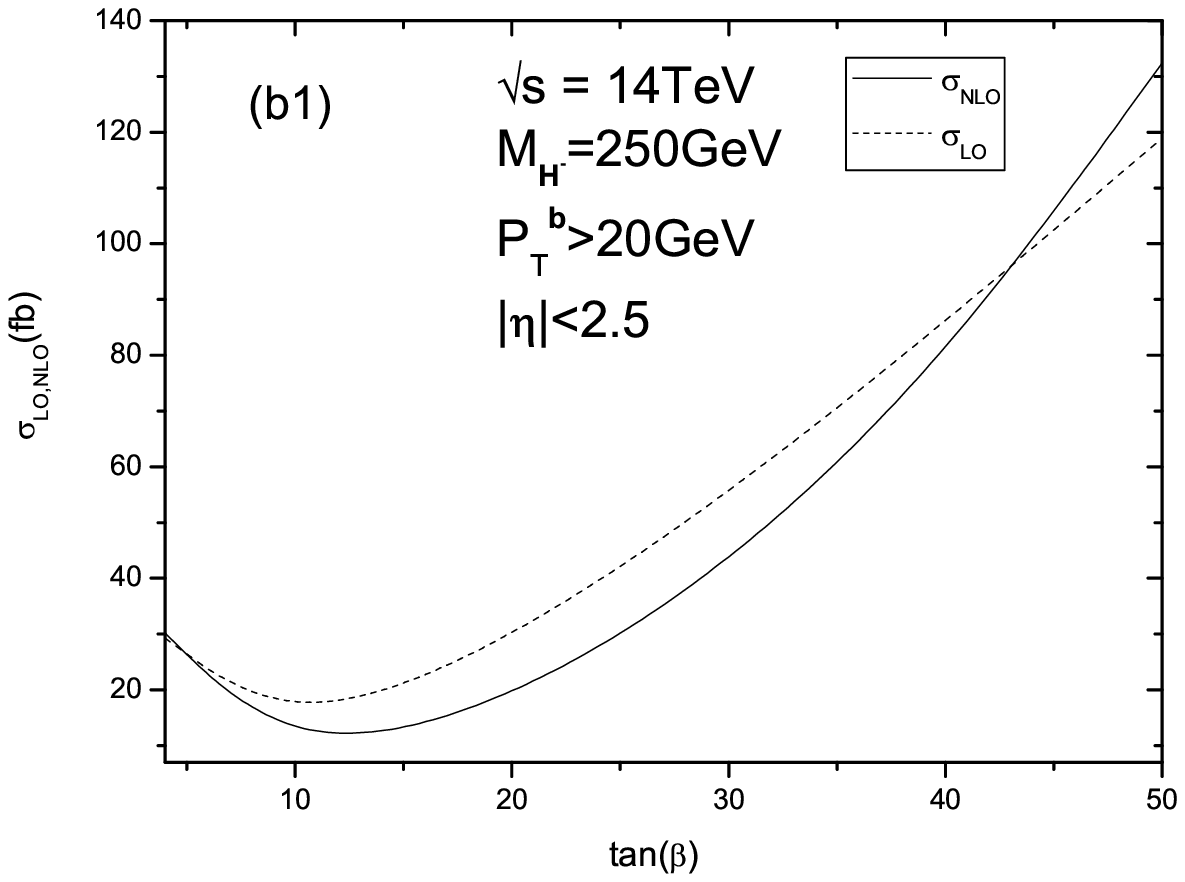}%
\hspace{0in}%
\includegraphics[width=3.2in,height=3in]{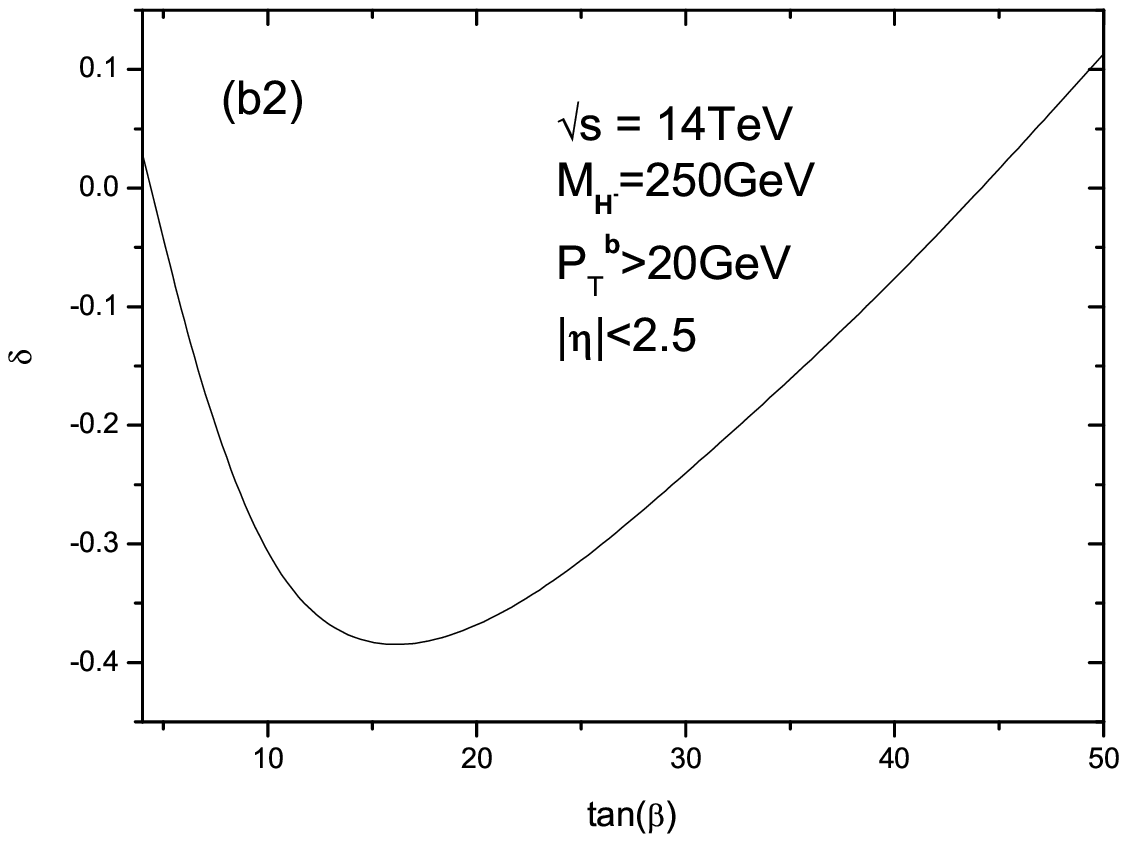}
\caption{The total NLO QCD corrected cross
sections($\sigma_{NLO}$) and the corresponding relative
corrections($\delta$) of the processes $p \bar{p}/pp \to t \bar
bH^-+X$ with $m_{H^-}=175~GeV$, $|\eta_b| \leq 2$ at the Tevatron
and $m_{H^-}=250~GeV$, $|\eta_b| \leq 2.5$ at the LHC, as the
functions of $\tan\beta$. Fig.8(a1) and Fig.8(a2) are for the
process $p\bar p \to t\bar bH^-+X$ at the Tevatron and Fig.8(b1)
and Fig.8(b2) for the process $pp \to t\bar bH^-+X$ at the LHC.}
\end{figure}

\begin{figure}[htp]
\centering
\includegraphics[width=3.2in,height=2.2in]{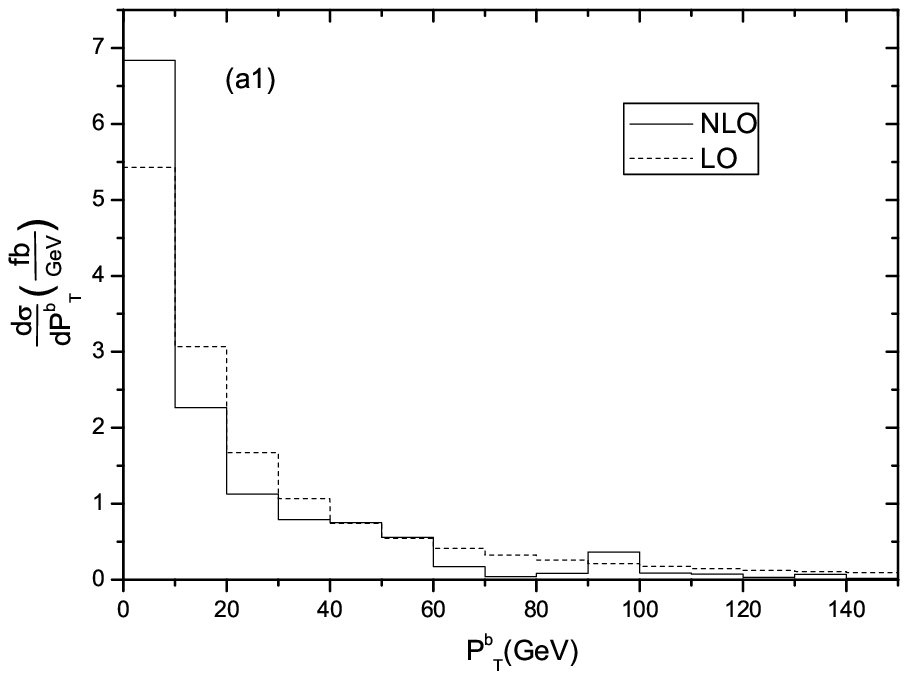}%
\hspace{0in}%
\includegraphics[width=3.2in,height=2.2in]{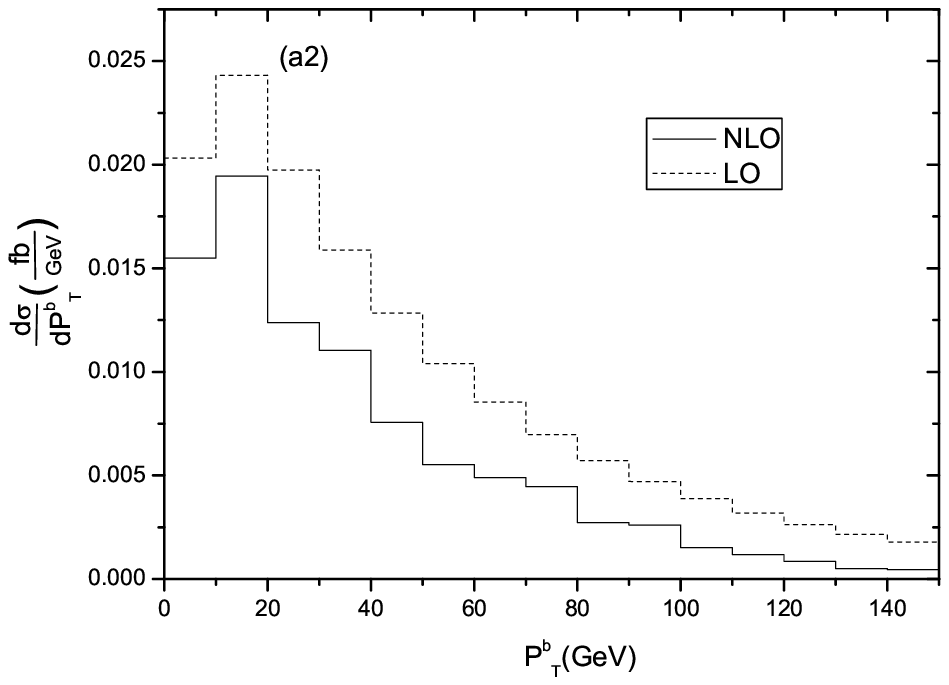}
\hspace{0in}%
\includegraphics[width=3.2in,height=2.2in]{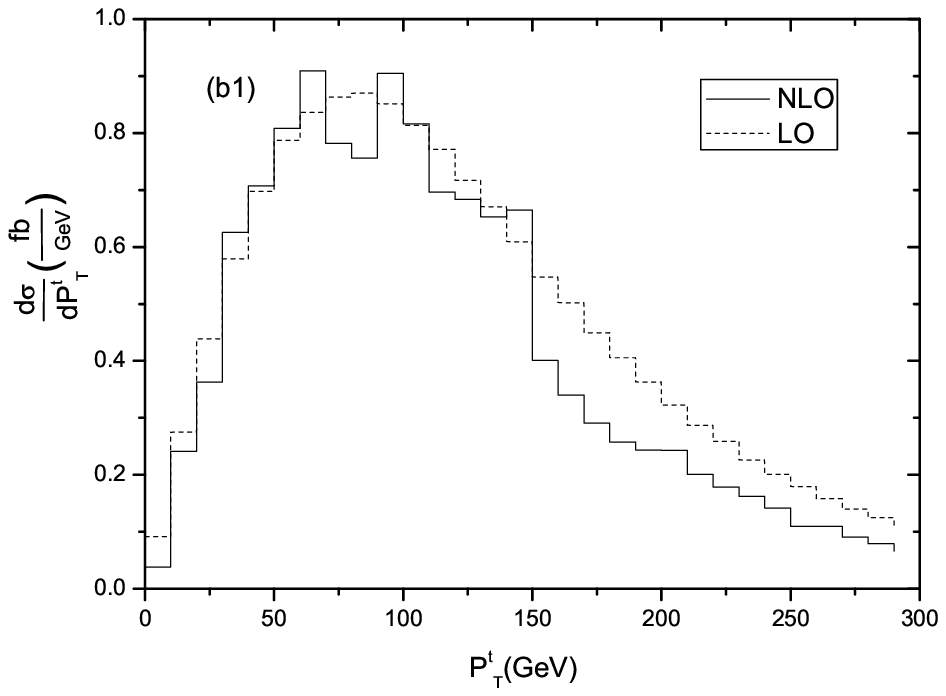}%
\hspace{0in}%
\includegraphics[width=3.2in,height=2.2in]{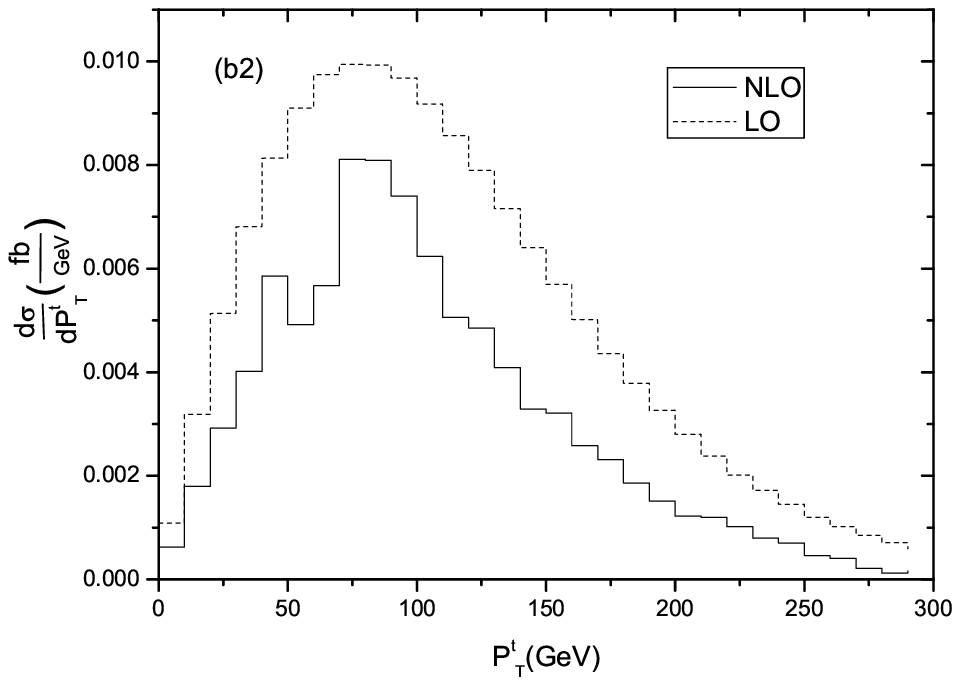}
\hspace{0in}%
\includegraphics[width=3.2in,height=2.2in]{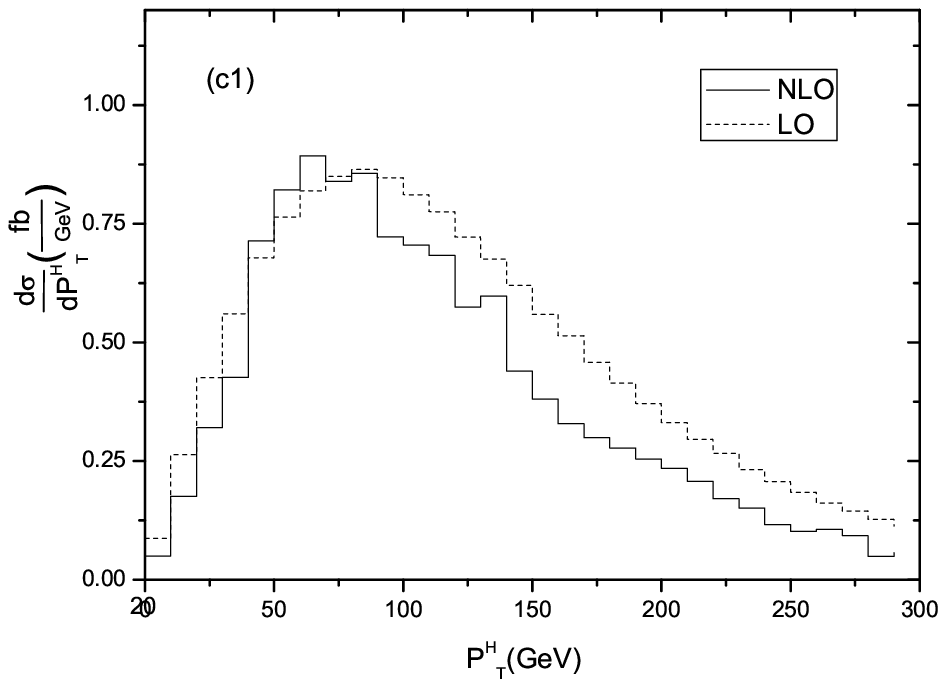}%
\hspace{0in}%
\includegraphics[width=3.2in,height=2.2in]{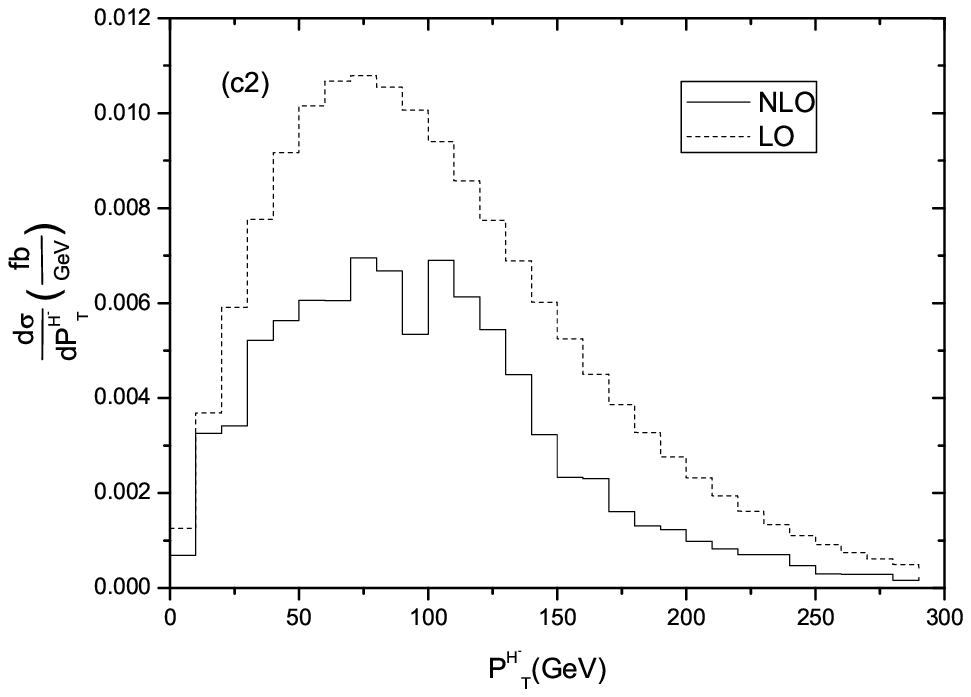}
\caption{The transverse momentum distributions of the final state
particles($t, \bar b, H^-$) at LO and NLO of the processes $p
\bar{p}/pp \to t \bar bH^-+X$ with $m_{H^-}=250~GeV$ at the LHC
and $m_{H^-}=175~GeV$ at the Tevatron. Fig.9(a1, b1, c1) are for
the transverse momentum distributions at the LHC and Fig.9(a2, b2,
c2) for the transverse momentum distributions at the Tevatron.}
\end{figure}

\end{document}